\documentclass{aa}
\usepackage{epsfig}
\usepackage{graphicx}
\usepackage{psfig,epsf}

\def\beq{\begin{equation}}
\def\eeq{\end{equation}}

\def\cm3{cm$^{-3}$}
\def\kms{km~s$^{-1}$}

\let\ts=\thinspace

\def\two{\ts {\sc ii}}

\def\beq{\begin{equation}}
\def\eeq{\end{equation}}

\begin{document}

\title{Quantitative spectroscopic analysis of and distance to SN1999em}
\subtitle{}

\author{Luc Dessart\inst{1,3}
        \and
        D. John Hillier\inst{2}
        }
\offprints{Luc Dessart,\\ \email{luc@as.arizona.edu}}

  \institute{Max-Planck-Institut f\"{u}r Astrophysik,
             Karl-Schwarzschild-Str 1, 85748 Garching, Germany 
        \and
           Department of Physics and Astronomy, University of Pittsburgh,
           3941 O'Hara Street, Pittsburgh, PA, 15260 
         \and
           Steward Observatory, University of Arizona, 933 North Cherry Avenue,
        Tucson, AZ 85721, USA
           }

\date{Accepted/Received}

\abstract{
Multi-epoch multi-wavelength spectroscopic observations of photospheric-phase
type II supernovae (SN) provide information on massive-star progenitor properties, 
the core-collapse mechanism, and distances in the Universe.
Following successes of recent endeavors (Dessart \& Hillier 2005ab)
with the non-LTE model atmosphere code CMFGEN (Hillier \& Miller 1998),
we present a detailed quantitative spectroscopic analysis of the type II SN1999em 
and, using the Expanding Photosphere Method (EPM) or synthetic fits 
to observed spectra, {\it \`a la} Baron et al. (2004),
we estimate its distance.
Selecting eight epochs, which cover the first 38 days after discovery, 
we obtain satisfactory fits to optical 
spectroscopic observations of SN1999em (including the UV and near-IR ranges when available).
We use the same iron-group metal content for the ejecta, the same power-law density distribution 
(with exponent $n = 10-12$), and a Hubble-velocity law at all times. We adopt a H/He/C/N/O 
abundance pattern compatible with CNO-cycle equilibrium values for a RSG/BSG progenitor, 
with C/O enhanced and N depleted at later times.
The overall evolution of the spectral energy distribution, whose peak shifts 
to longer wavelengths as time progresses, reflects the steady  
temperature/ionization-level decrease of the ejecta, associated non-linearly
with a dramatic shift to ions with stronger line-blocking powers in the UV and optical 
(Fe{\,\sc ii}, Ti{\sc ii}).
In the parameter space investigated, CMFGEN is very sensitive and provides photospheric 
temperatures and velocities, reddenings, and the H/He abundance ratio with an accuracy 
of $\pm$ 500\,K, $\pm$ 10\%, 0.05 and 50\%, respectively.
Following Leonard et al. (2002), and their use of correction factors from Hamuy et al. (2001), we estimate 
an EPM distance to SN1999em that also falls 30\% short of the Cepheid distance of 11.7\,Mpc
to its host galaxy NGC\,1637 (Leonard et al. 2003). However,
using the systematically higher correction factors of Dessart \& Hillier (2005b) removes the discrepancy.
A significant scatter, arising primarily from errors in the correction factors and derived temperatures,
is seen in distances derived using different band passes. 
However, adopting {\it both correction
factors and corresponding color-temperatures} from tailored models to each observation leads 
to a good agreement between distance estimates obtained from different band passes.
The need for detailed model computations thus defeats 
the appeal and simplicity of the original EPM method, which uses
tabulated correction factors and broadband fluxes, for distance determinations.
However, detailed fits to SN optical spectra, based on tailored models for individual SN observations, 
offers a promising approach to obtaining accurate distances, either through the EPM or via the
technique of 
Baron et al. (2004). Our best distance-estimate to SN1999em is $11.5 \pm 1.0$\,Mpc.
We note that to achieve 10-20\% accuracy in such distance estimates requires multiple
observations, covering preferentially a range of early epochs preceding the hydrogen-recombination phase. 

}
\titlerunning{Type II supernovae spectroscopic modeling}
\maketitle

\section{Introduction}
\label{Sec_intro}

Stellar explosions, due to their large luminosity, represent an excellent
probe of the distant, and thus young, Universe.
Short Gamma Ray Bursts, identified with the death of Wolf-Rayet stars
(MacFadyen \& Woosley 1999; MacFadyen et al. 2001), rank top of the prospective 
list, but currently, these phenomena are not suitably understood for distance 
calibration purposes.
Down on the energetics ladder, supernovae explosions, discovered as bursts of
optical light, have thus far been a viable route.
The inherent uniformity of Chandrasekhar-mass white-dwarfs, associated with type Ia
SN, has motivated their nomination as standard candles.
For the accuracy level sought, however, dust obscuration, metallicity effects, and (unknown)
progenitor differences might invalidate such an assumption.
Although typically ten times less luminous than type Ia, type II SN, resulting
from the collapse of the Fe-core of a massive star, represent an alternative.
By assuming their ejecta expand homologously and radiate a blackbody spectrum
at a well-defined temperature, Kirshner \& Kwan (1974) designed the Expanding 
Photosphere Method (EPM), a variant of the Baade (1926) method for variable stars, 
to constrain the distance to a SN.
The agreement between the EPM-distance to SN1987A and that obtained, e.g., with
LMC Cepheids, warranted application to a bigger sample, stretching into the Hubble
flow and thus providing some independent constraint on the Hubble constant
(Schmidt et al. 1992).
But the 30-50\% lower value of the EPM-distance to SN1999em (Leonard et al. 2002, L02;
Hamuy et al. 2001, H01) compared to the subsequently obtained Cepheid-distance to its
host galaxy (NGC\,1637; Leonard et al. 2003) led to question the reliability of the EPM.
Using the Spectral-fitting Expanding Atmosphere Method (SEAM), Baron et al. (2004, B04)
found a good agreement between their inferred distance to SN1999em and the Cepheid-distance
to NGC\,1637, and suggested that correction factors, used to approximate the
Spectral Energy Distribution (SED) with that of a single-temperature blackbody, 
were too small at later epochs.

In this paper, we study the various ingredients entering the EPM to hunt for the origin 
of the above discrepancy, surprising because both methods share a number of assumptions
or uncertainties: they both assume homologous expansion, are both affected by uncertainties 
in outflow expansion velocity or photometric measurements, and rely on multi-epoch observations 
to constrain the poorly-known time of explosion.
Besides the generic 20\% scatter of correction factors, the 20-50\% lower values found by 
Eastman et al. (1996, E96)  compared to those of Dessart \& Hillier (2005b, Paper II) 
could partly or fully explain the underestimate of the distance to NGC\,1637 and SN1999em 
based on the EPM (H01, L02).
Our approach to determining distances to type II SN requires a detailed modeling of multi-epoch 
spectroscopic observations; we thus also infer progenitor and SN outflow properties.


In Dessart \& Hillier (2005a, Paper I), we presented promising first analyses of
photospheric-phase type II SN with CMFGEN (Hillier \& Miller 1998), a model atmosphere code
so far devoted to studies of massive star winds (Najarro et al. 1997, Hillier \& Miller 1999,
Dessart et al. 2000, Crowther et al. 2002, Martins et al. 2004).
To step progressively through the various challenges posed by SN spectroscopic
modeling, we first concentrate on type II SN, known to exhibit modest,
and at early times negligible, amounts of nucleosynthesised metals during the
explosion, thereby allowing us to neglect the deposition of
energy from radioactive decay.
Together with a Hubble velocity law, we implemented a simple power law
density distribution $\rho(r) = \rho_0 (R_0/r)^n$, where $\rho_0$ and $R_0$ are
the base density and radius, and $n$ is the density exponent.
Using spectroscopic observations of SN1987A and SN1999em, selected at times when hydrogen
is partially or fully ionized, we found that both objects could be well modeled
by adopting standard CNO-cycle equilibrium abundances typical of a supergiant massive
star progenitor. Two optical features in early-time spectra, previously associated with 
peculiar line emission/absorption processes and/or outflow inhomogeneities lying high 
above the photosphere (Baron et al. 2000), were unambiguously identified as N{\,\sc ii}-multiplet 
lines. The strength of the features suggested a nitrogen enrichment, at the corresponding epoch, 
of a factor of a few compared to cosmic for SN1999em, but somewhat less for SN1987A. 
He{\,\sc i}\,5875\AA\, could also be well
reproduced with a modest helium enrichment, in contrast with former difficulties associated with 
that species (Eastman \& Kirshner 1989). 
Additionally, optical line profiles, and in particular their strong and persistent blueshift,
were found to place a strong constraint on the density-gradient
in the vicinity of the photosphere, requiring $n \sim 10$.
For these test cases, CMFGEN fitted the UV, the optical and the near-IR (when available)
with an accuracy generally better than a few tens of percent. 

Overall, this high-level of agreement between our synthetic SEDs and spectroscopic observations
of type II SN during their photospheric phase validates such a detailed approach with CMFGEN.
It allows us to draw conclusions on the progenitor properties and perhaps more importantly 
on distances to the host galaxy.

In Dessart \& Hillier (2005b), we reviewed the various items entering the
EPM, to complement the only other study of the kind done by Eastman et al. (1996, E96).
CMFGEN yields correction factors that are systematically $\sim$20\% larger than those
of E96, but confirms the general properties with effective- or
blackbody color-temperature, and photospheric density.
We also stressed that correction factors do not only correct for dilution by electron
scattering of the SED set at the thermalization depth, but, within the standard domain of
application of the EPM, also correct for the corrupting effects of line
emission and absorption on the continuum flux within a given bandpass.
Following the appearance of metal line blanketing below $T_{\rm phot} \sim 8000$K,
such line effects dominate in the $B$-band (attributable to Fe{\,\sc ii} and Ti{\,\sc ii})
and in the $I$-band (Ca{\,\sc ii}), making the SED incompatible with that of a blackbody.
Due to the steep increase of correction factors at low color-temperatures,
modulations of line-strengths (having nothing to do with ionization but resulting
from different metal abundances or expansion velocity) can affect inferred
color-temperatures and generate substantial errors.
Moreover, associated optical-depth enhancements in the corresponding spectral regions
make the photosphere more extended than in the nearby line-free regions,
invalidating the uniqueness of the photosphere radius required by the Baade method.
These conclusions thus argue for the use of the EPM at early-times, when the outflow 
is fully ionized and metal-line blanketing confined to the UV region.
Finally, although the velocity at maximum absorption $v_{\rm max}$ in optical P-Cygni profiles
corresponds to the photospheric velocity $v_{\rm phot}$ to within few tens of percents, we
showed that $v_{\rm max}$ can over- as well as under-estimate $v_{\rm phot}$, i.e.  
$v_{\rm max}$ {\it does not necessarily} exceed $v_{\rm phot}$ (in absolute terms).

In this paper, we analyze a high-quality photometric and spectroscopic dataset 
for SN1999em with CMFGEN and infer its distance, first with the EPM in combination with 
various sources for correction factors, and then using CMFGEN fits to each epoch 
(an approach similar in spirit to the SEAM of B04).
In Sect.~\ref{Sec_obs}, we present our selection of spectro-photometric observations,
according to the criteria outlined above.
In Sect.~\ref{Sec_mod}, we show the results of the quantitative spectroscopic modeling
for each of the selected dates, drawing information on the outflow kinematics,
chemistry, ionization.
In Sect.~\ref{Sec_dist}, we compute the distance to SN1999em.
In Sect.~\ref{Sec_epm}, using the standard minimization procedure and prescriptions 
for correction factors from H01, the EPM-distance 
shows a large scatter between selected band passes, is comparable to that found by L02 
and H01 and thus always lower than the Cepheid-distance to NGC\,1637. When
we use the correction factors from Paper II, distance estimates are in better
agreement with the Cepheid-distance, but there is still a relatively large
scatter between different filter sets. A similar scatter occurs when we  
use blackbody color-temperatures extracted from the detailed models with the
correction factors tabulated in Paper II. As expected, consistent distance estimates
are obtained when we use the model blackbody color-temperatures together with the model
correction factors.
For comparison, we compute the SEAM-distance to SN1999em in Sect.~\ref{Sec_seam}
and corroborate the EPM-distance determination based on CMFGEN models or 
that of B04.
In Sect.~\ref{Sec_conc}, we draw our conclusions and discuss future prospects.

\section{Spectrophotometric datasets of and reddening to SN1999em}
\label{Sec_obs}

No neutrino detection was associated with the SN1999em event and thus no precise dating of
the core-collapse of the progenitor exists.
Inferences, based on the light curve appearance or the EPM/SEAM, suggest
the explosion took place around the 25th of October 1999, with an uncertainty of $\pm$ 2 days
(H01, L02, B04).
It was absent in images of the host galaxy NGC\,1637 taken on October 20.45 UT,
while first detection of the SN was made on October 29th (JD 2451480.94; Li 1999).
The resemblance of the spectrum taken on October 30th with the spectrum of SN1987A
taken on the 24th of February 1987, indicates that SN1999em was indeed caught at a very early
phase of evolution.

In this work, we use two main sources of spectroscopic observations to model the
evolution of SN1999em, from the first spectrum taken on the 30th of October 1999
until the last one taken on the 5th of December 1999, by which time metal line-blanketing
in the optical spectral range is strong.
From H01, we use optical spectra taken on the 30th of October, and the 3rd, 9th, 14th,
and 19th of November.
In addition, we use the near-IR observations of H01 on the 2nd and 19th of
November.
From the SUSPECT\footnote{www.bruford.nhn.ou.edu/\~\,suspect} archive, we extract
optical data of L02 for the 1st and 5th of November, and the 5th of December.
Finally, we use HST--STIS observations taken on the 5th of November, kindly
provided by Baron (priv. comm.), and covering the wavelength range from 1140\AA\,
to 5710\AA, overlapping in the red with observations of L02. On that date, we merge
the two datasets to have a full coverage of the UV and optical out to 7500\AA.
We exclude from our sample the observations of the 22nd and 24th of November (L02)
because of dubious absolute flux level, curious relative flux distribution at and
beyond H$\alpha$ (compared to spectra taken at embracing dates; see discussion in L02).
Hence, our sample contains eight observations for which photometric measurements have
been obtained by L02, coincident to within a few hours to spectroscopic ones.
On day 38, we adopt the magnitudes of L02 and combine these measurements with the spectrum
taken on the previous day, i.e., the 5th of December, justified by the very slow 
spectroscopic evolution at such late times.
%
In Table~\ref{tab_obs}, we provide a log of the observations used in this work.
Despite H01's photometric measurements over a large time-span in the $Z$, $J$,
$H$, and $K$ bands, we limit our investigation to the optical band passes $B$, $V$, 
and $I$, and sets $\{B,V\}$, $\{B,V,I\}$, and $\{V,I\}$ (note one exception with 
the $Z$-band in Sect.~\ref{Sec_seam}; see H01 for a presentation of this $Z$-band).


Our modeling procedure blueshifts the spectroscopic observations of SN1999em by
770\,km\,s$^{-1}$ (L02), scales the synthetic flux to the Cepheid-distance of 11.7\,Mpc (L03), and
finally reddens such a scaled synthetic spectrum using the Cardelli et al. (1998) law,
adopting $A_V/E(B-V)=3.1$.
In Paper I, we illustrated the mingled effects on the SED associated with metal
line-blanketing, cooling of the outflow and reddening.
Here, the fine time-sampling and the HST-STIS data can only be fitted with CMFGEN by adopting
a low reddening value of 0.1$\pm$0.05, in agreement with previous studies (L02, Baron et al. 2000, 
B04; see also Sect.~\ref{Sec_teff_red}).
Note that the UV data is indeed a key asset for reddening determinations since the SED at
longer wavelength becomes less and less sensitive to reddening modulations.
In the context of the EPM, the extinction uncertainty translates directly into
an uncertainty in the distance modulus. Fortunately, due to the low ISM column density to
SN1999em, this introduces modest absolute errors on the distance
(see Sect.~\ref{Sec_seam}).

To highlight line-emission contributions, we always plot the synthetic spectrum
obtained by including all opacity sources (red curve) and that accounting only for 
continuum processes (blue curve). 
Finally, the adopted model luminosities for each date match the absolute observed flux
level (for the assumed Cepheid-distance) to within a factor of approximately 2-3 (our  
initial adoption of the lower distance estimate by H01/L02 when modelling SN1999em introduced a systematic
underestimate of the luminosities by a factor of $\sim$(11.7/8)$^2 \sim$2.1);
the synthetic SED shows only a modest sensitivity to luminosity scalings of that magnitude, 
provided the outflow ionization is preserved (Paper I, E96).


In Sects.~\ref{Sec_epm}-\ref{Sec_seam}, we convert synthetic and blackbody fluxes
into magnitudes using the filter transmission functions and zero-points
of H01 and Hamuy (priv. comm.), adopting either the value at each filter's effective
wavelength
($A_B = 4.05\,E(B-V)$
and $A_I =  1.83\,E(B-V)$),
or convolving the synthetic SED with the filter transmission function.
Resulting magnitudes from the two approaches differ by no more than 0.005 mag,
thus much less than other uncertainties involved in the EPM or SEAM.


\begin{table}
\caption[]{Log of the observations used and described in Sect.~\ref{Sec_obs}.
Photometric measurements, with corresponding dates, are reproduced from L02
(columns 2--5). Spectroscopic datasets are referenced according to spectral range,
i.e. UV, optical (Opt.) and near-IR (nIR). Note that in Sect.~\ref{Sec_dist}, the photometric
data on day 38 is combined with the spectrocopic data on day 37.}
\label{tab_obs}
\begin{tabular}{lc|ccc|ccc}
\hline
Date  & Day$^a$ & \multicolumn{3}{|c|}{Photometry$^b$} & \multicolumn{3}{|c}{Spectroscopy} \\
\hline
  (1999)      &     &   $B$  & $V$   & $I$    &      UV   &   Opt.  & nIR  \\
\hline
30 Oct & 1.0 & 13.87  & 13.87 & 13.65  &    ...    &   1      &  ...      \\
01 Nov & 3.0 & 13.80  & 13.79 & 13.56  &    ...    &   2      &  ...      \\
02 Nov & 4.0 & ...    & ...   & ...    &    ...    &   ...    &   1       \\
03 Nov & 5.0 & 13.85  & 13.79 & 13.54  &    ...    &   1      &  ...      \\
05 Nov & 7.0 & 13.92  & 13.84 & 13.53  &    3      &   2,3    &  ...      \\
09 Nov & 11.0& 14.02  & 13.84 & 13.48  &    ...    &   1      &  ...      \\
14 Nov & 16.0& 14.25  & 13.81 & 13.44  &    ...    &   1      &  ...      \\
19 Nov & 21.0& 14.47  & 13.86 & 13.40  &    ...    &   1      &   1       \\
06 Dec & 38.0& 14.94  & 13.93 & 13.29  &    ...    &   2      &  ...      \\
\hline\\
\end{tabular}\\
\begin{flushleft}
a: days since JD 2,451,480.94; b: measurements are taken from 1: H01; 2: L02; 3: Baron et al. (2000).
\end{flushleft}
\end{table}

\section{Quantitative spectroscopic analysis of SN1999em}
\label{Sec_mod}

In this section, we perform a quantitative spectroscopic analysis of
SN1999em, at eight dates sampling its early photospheric-phase evolution (see above).
In Sect.~\ref{Sec_mod_pres}, we review the standard model parameters and assumptions.
We then detail our results for each observation in Sect.~\ref{Sec_mod_analysis}.
Finally, in Sect.~\ref{Sec_mod_disc}, we discuss typical uncertainties associated with
each of the inferred model parameters.

\subsection{Model Presentation}
\label{Sec_mod_pres}

\begin{table*}
\begin{center}
\caption[]{Summary of model results presented in Sect.~\ref{Sec_mod_analysis}.
Time reference is in days since JD 2,451,480.94. 
On the right hand side, we also give the key EPM quantities for each model and date, 
i.e., blackbody color temperature and corresponding correction factor for each 
bandpass, $\{B,V\}$, $\{B,V,I\}$, and $\{V,I\}$ (Sect.~\ref{Sec_dist}; Paper II).
}
\label{tab_mod_analysis}
\begin{tabular}{l|cccccc|cccccc}
\hline
Day  & $L_{\ast}$ &$T_{\rm phot}$ & $R_{\rm phot}$&$v_{\rm phot}$&$\rho_{\rm phot}$&$n$&
$T_{BV}$ & $\xi_{BV}$  &$T_{BVI}$ & $\xi_{BVI}$  &$T_{VI}$ & $\xi_{VI}$   \\
\hline
         & (10$^8$ $L_{\odot}$) & (kK) &  (10$^{14}$\,cm) &   (\kms)   & (10$^{-14}$\,g\,cm$^{-3}$)&
& (kK)&  & (kK)&  & (kK) &    \\
\hline
1.0      & 15.0  & 13.38 & 6.2  & 10,430  & 4.3  & 12 &16.02 & 0.462 & 14.96 & 0.50 &14.39&  0.52 \\
3.0      & 15.0  & 12.59 & 6.8  &  8,940  & 3.3  & 10 &15.96 & 0.430 & 15.03 & 0.46 &14.23&  0.49 \\
4.0      & 15.0  & 12.59 & 6.8  &  8,940  & 3.3  & 10 &15.96 & 0.430 & 15.03 & 0.46 &14.23&  0.49 \\
5.0      &  9.0  & 10.63 & 6.7  &  8,840  & 3.6  & 10 &14.23 & 0.443 & 13.20 & 0.49 &12.53&  0.51 \\
7.0      &  5.0  &  9.20 & 6.64 &  8,750  & 4.1  & 10 &12.95 & 0.443 & 11.95 & 0.49 &11.28&  0.53 \\
11.0     &  2.5  &  8.04 & 6.65 &  7,960  & 4.0  & 10 &10.70 & 0.469 & 10.25 & 0.50 & 9.83&  0.53 \\
16.0     &  1.5  &  6.80 & 6.15 &  6,350  & 8.7  & 10 & 8.17 & 0.646 &  8.84 & 0.56 & 9.55&  0.51 \\
21.0     &  3.75 &  6.26 &12.90 &  5,530  & 7.4  & 10 & 7.11 & 0.685 &  7.85 & 0.56 & 8.74&  0.48 \\
38.0     &  2.0  &  5.92 &10.90 &  3,400  &38.8  & 10 & 5.54 & 1.072 &  6.34 & 0.75 & 7.24&  0.61 \\
\hline
\end{tabular}
\end{center}
\end{table*}

The standard set of parameters needed by CMFGEN were discussed thoroughly in Paper I.
A key asset of CMFGEN is the explicit treatment of line-blanketing, justifying a detailed
description of the atomic structure for a large number of species.
Models discussed below include H{\,\sc i} (30,20), He{\,\sc i} (51,40), C{\,\sc i} (63,33),
C{\,\sc ii} (59,32), C{\,\sc iii} (20,12), C{\,\sc iv} (14,9), N{\,\sc i} (104,44),
N{\,\sc ii} (41,23), N{\,\sc iii} (8,8), O{\,\sc i} (75,23), O{\,\sc ii} (111,30),
O{\,\sc iii} (46,26), Ne{\,\sc ii} (242,42), Na{\,\sc i} (71,22), Na{\,\sc ii} (35,21),
Mg{\,\sc ii} (65,22), Al{\,\sc ii} (44,26), Al{\,\sc iii} (45,17), Si{\,\sc ii} (59,31),
Si{\,\sc iii} (51,27), S{\,\sc ii} (324,56), S{\,\sc iii} (98,48), Ca{\,\sc ii} (77,21),
Ti{\,\sc ii} (152,37), Ti{\,\sc iii} (206,33), Cr{\,\sc ii} (196,28), Cr{\,\sc iii} (145,30),
Mn{\,\sc ii} (97,25), Mn{\,\sc iii} (175,30), Fe{\,\sc i} (136,44), 
Fe{\,\sc ii} (309,116), Fe{\,\sc iii} (477,61),
Fe{\,\sc iv} (282,50), Co{\,\sc ii} (144,34), Co{\,\sc iii} (283,41), Ni{\,\sc ii} (93,19),
Ni{\,\sc iii} (67,15), each parenthesis containing the number of full- and super-levels
(see Hillier \& Miller 1998 for details).
At epochs when a given species offers a negligible opacity to the radiation field at all
depths in the outflow, we exclude it from the atomic dataset, thereby reducing the
memory requirements and accelerating the convergence of CMFGEN.
Under typical type II SN conditions and when all the above species are included, CMFGEN uses
up to 3.5GB of RAM and may take a week to reduce all level population fluctuations,
from one iteration to the next, to less than 1\%.

As discussed below, the main component controlling the evolution of type II SN
spectra during their photospheric phase is the steady decrease of the outflow
temperature, in parallel with the reduction of the outflow ionization.
As time proceeds, opacity sources change, and the opacity may increase
(e.g., when Fe{\,\sc ii} dominates over Fe{\,\sc iii}).
In contrast to the large changes in effective temperature,
we find only small and specific changes in chemical composition as time progresses 
through the first 40 days past discovery. 
No metal-abundance variations associated with explosive nucleosynthesis are found,
so that all spectra are fitted with a unique metallicity, adopted to be solar (L02).
For hydrogen and helium, we reproduce all epochs with a unique value, H/He = 5 
(by number), although our inability to fit the Balmer lines at late times makes 
our result reliable only up to the hydrogen-recombination phase (first twenty days).
The nitrogen abundance is constrained from the N{\,\sc ii} optical lines (Paper I) 
and requires over-solar abundance; we adopt N/He = 6.8 $\times 10^{-3}$ 
(Prantzos et al. 1986).
C and O abundances are difficult to constrain: at early times, since no lines 
from these species are observed, we adopt C and O abundances compatible
with the N/He above and CNO-cycle equilibrium values, with C/He = 1.7 $\times 10^{-4}$ and
O/He = 10$^{-4}$ by number.
At later times, however, C{\,\sc i} and O{\,\sc i} lines are unambiguously observed
in the 7000-12\,000\AA\, region, requiring enhancements compared to CNO equilibrium
values, of a factor of 10 for carbon and 100 for oxygen (Sect.~\ref{Sec_0512}).
Finally, the Na{\,\sc i}\,5890\AA\, doublet can only be fitted with a factor of four 
over-abundance compared to solar, in agreement with the CNO-cycle equilibrium values
of Prantzos et al. (1986), i.e. $X_{\rm Na}$ = 1.38 $\times$ 10$^{-4}$.
Overall, we note that the conspicuous spectroscopic evolution during the 
photospheric phase is not a reflection of outflow changes in chemical composition 
but rather in ionization/temperature.

%
%

The density distribution of the outflow is constrained {\it via} the density
exponent $n$.
In Papers I and II, we illustrated, from different perspectives, the spatial confinement
of line and continuum formation regions to a few density scale heights about the
photosphere rather than over the entire outflow - the Ca{\,\sc ii} set of lines at 
around 8500\AA\, is a noticeable exception (e.g., Fig.~13., Paper I).
The resulting line-profile shapes require values $n \ga 10$ to reproduce the blueshift
as well as both the modest absorption and emission observed in Balmer lines.
Comparable values are obtained in other spectroscopic studies, e.g. Eastman \& Kirshner (1989),
or independent radiation-hydrodynamical simulations of core-collapse (Arnett 1988;
Woosley 1988; Ensman \& Burrows 1992).

We quote numerous model properties at the photosphere, which corresponds here to the location 
where the continuum optical depth, integrated inwards from the outer grid radius, is 2/3.
We constrain the expansion, or, equivalently, the photospheric velocity by adjusting 
the velocity at the inner boundary;
we then adopt the value that permits a satisfactory fit to profile shapes for 
a wide range of optical line diagnostics, a procedure we favor over the direct and 
more precise, but less meaningful, measurements on individual observed line profiles.
Further discussion on this is provided in Sect. 5 of Paper II, as well as below, in
Sect.~\ref{Sec_mod_disc}.

\subsection{Tailored Analysis with CMFGEN}
\label{Sec_mod_analysis}

   In this section we model the data for our selected observation dates (Sect.~\ref{Sec_obs}),
provide synthetic fits to observed SEDs (Figs.~\ref{fig_3010}-\ref{fig_0512}), and tabulate the derived
model parameters (Table~\ref{tab_mod_analysis}).
As described in the preceding section, the outflow temperature, modulated via
the base luminosity or radius, represents the essential change between consecutive dates.
The electron temperature at the photosphere, $T_{\rm phot}$, accounts well for this change, and
thus constitutes the center of the discussion below. To infer the photospheric
velocity, many models were calculated: we adopted the photospheric
velocity corresponding to that model which provided the best fit to
the majority of observed line profiles. A detailed discussion on the uncertainties in the 
derived photospheric velocity is provided in Sect.~\ref{Sec_phot_vel}.


\subsubsection{Observation of the 30th of October 1999}
\label{Sec_3010}

\begin{figure}
 \epsfig{file=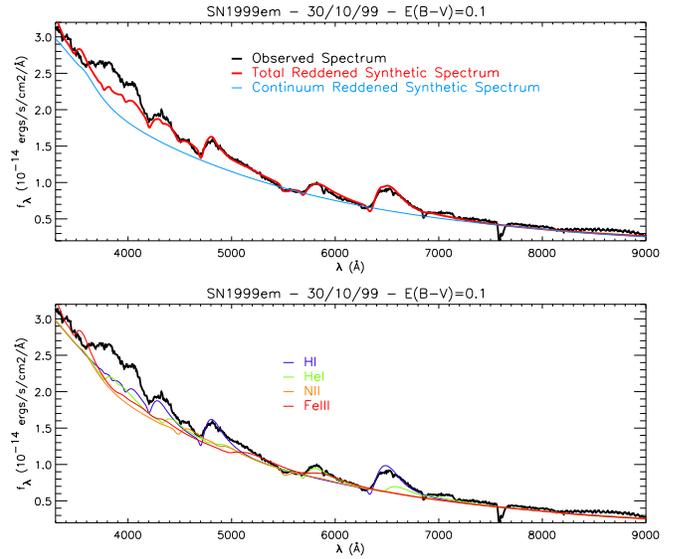, width=9cm}

\caption{
{\it Top}: fit of the full (red) and continuum (blue) synthetic spectra
to observations (black) of SN1999em taken on the 30th of October 1999 (Day 1; H01).
Model parameters are: $L_{\ast} = 1.5 \times 10^9 L_{\odot}$,
$T_{\rm phot} = 13\,380$K, $R_{\rm phot}=6.2 \times 10^{14}$\,cm,
$v_{\rm phot}=$10\,430\,km\,s$^{-1}$, $\rho_{\rm phot} = 4.3 \times 10^{-14}$\,g\,cm$^{-3}$,
and $n=12$.
The synthetic flux, 17\% lower than observed, is re-normalized at 8200\AA.
{\it Bottom}: same as above, but this time overplotting reddened synthetic spectra 
computed by treating bound-bound transitions of a {\it single} species: H{\,\sc i}, 
He{\,\,\sc i}, N{\,\sc ii}, and Fe{\,\sc iii}.
Contrary to Baron et al. (2000), we attribute the two features just blue-ward of H$\beta$ and
He{\,\sc i}\,5875\AA\, to N{\,\sc ii} lines. 
(This figure is available in color in the electronic version.)
}
\label{fig_3010}
\end{figure}

In the top panel of Fig.~\ref{fig_3010}, we present a fit of the reddened scaled
synthetic spectrum (full: thick red; continuum: thin blue) to the observations
taken on the 30th of October 1999 (black; H01).
The model parameters are:  $L_{\ast} = 1.5 \times 10^9 L_{\odot}$,
$T_{\rm phot} = 13\,380$\,K, $R_{\rm phot}=6.2 \times
10^{14}$\,cm, $v_{\rm phot}=10\,430$\,km\,s$^{-1}$, $\rho_{\rm phot} = 4.3 \times
10^{-14}$\,g\,cm$^{-3}$, and $n=12$.
The flux, underestimated by 17\%, is renormalised at 8200\AA.
The overall fit quality is good, although the flux around 3800\AA\, is underestimated.
Including more species or bigger model atoms makes no difference; increasing $T_{\rm phot}$
helps but then provides a poor match to the observed flux elsewhere, e.g., below 3500\AA.
A poor flux calibration in this spectral region could be at the origin of this discrepancy.
The fit quality to Balmer lines is satisfactory, although
the extent of profile troughs is underestimated.
At such early times, our neglect of relativistic terms in the radiative transfer equation
may contribute to such discrepancies (Hauschildt et al. 1991; Jeffery 1993).
Helium lines, and most notably He{\,\sc i}\,5875\AA, are well reproduced.
In the bottom panel of Fig.~\ref{fig_3010}, we show synthetic spectrum fits (thin colored-lines)
to observations, but treating chemical species in isolation.
As emphasized in Paper I, we predict N{\,\sc ii} (orange line) features in the blue wing of
He{\,\sc i}\,5875\AA, resulting from the 3d-3p multiplet  around 5470\AA\, and
the 3p-3s multiplet around 5670\AA, as well as in the blue wing of H$\beta$, resulting
from other N{\,\sc ii} multiplets around 4600\AA.
Finally, notice how the continuum lies in general well below the full synthetic flux
distribution level, even in spectral regions that look essentially featureless, 
revealing the ubiquitous presence of line emission (see, e.g., the subtle contribution 
from Fe{\,\sc iii} -- colored in red --  to the red wing of both H$\beta$ and He{\,\sc i}\,5875\AA,
or that of He{\,\sc i}\,6678\AA\, in the red wing of H$\alpha$).

\subsubsection{Observations of the 1st--2nd of November 1999}
\label{Sec_0111}

\begin{figure}[htp!]
\epsfig{file=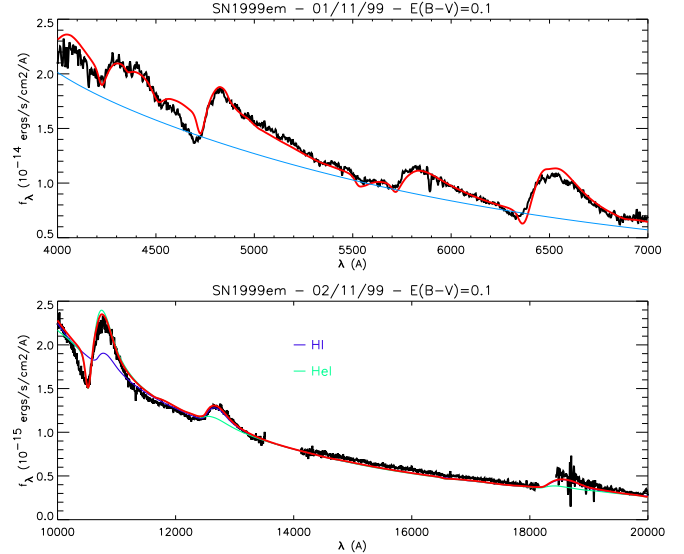, width=9cm}
\caption{
Fit of the full (red) and continuum (blue) synthetic spectra to observations (black)
of SN1999em taken on the 1st (optical, L02, day 3; top) and the 2nd (near-IR, H01,
day 4; bottom) of November 1999.
In the bottom panel, we include the synthetic spectra obtained by restricting bound-bound transitions
to those of H{\,\sc i} (blue) or He{\,\sc i} (green).
The model parameters are: $L_{\ast} = 1.5 \times 10^9 L_{\odot}$, $T_{\rm phot} = 12\,590$\,K,
$R_{\rm phot} = 6.8 \times 10^{14}$\,cm, $v_{\rm phot}=8940$\,km\,s$^{-1}$,
$\rho_{\rm phot} = 3.3 \times 10^{-14}$\,g\,cm$^{-3}$ and $n=10$.
The model flux, which underestimates the observations by 30\%, 
is renormalized to the observed value at 6900\AA.
(This figure is available in color in the electronic version.)
}
\label{fig_0111}
\end{figure}

In Fig.~\ref{fig_0111}, we reproduce Fig.~\ref{fig_3010} for the observations of
the 1st (optical, top panel; L02) and the 2nd (near-IR, bottom panel; H01) of November 1999.
The full (thick red) and continuum (thin blue) synthetic flux distributions underestimate
by 30\% the observed flux and are thus re-normalized, at 6900\AA, to fit both
spectral ranges (with an additional 8\% flux adjustment between optical and near-IR).
The model parameters, identical for both spectral ranges, are:
$L_{\ast} = 1.5 \times 10^9 L_{\odot}$, $T_{\rm phot} = 12\,590$\,K,
$R_{\rm phot} = 6.8 \times 10^{14}$\,cm, $v_{\rm phot} = 8940$\,km\,s$^{-1}$,
$\rho_{\rm phot} = 3.3 \times 10^{-14}$\,g\,cm$^{-3}$, and $n=10$.

In the optical, there is little change with observations taken on the 30th of October,
reflected in the similarity of model parameters for both dates.
We obtain good fits to Balmer as well as Paschen lines, i.e.,
P$\gamma$\,1.094\,$\mu$m, P$\beta$\,1.28\,$\mu$m, and P$\alpha$\,1.875\,$\mu$m.
Note, however, that profile troughs are sometimes predicted too narrow (e.g., H$\alpha$
and H$\beta$).
He{\,\sc i} lines are also well reproduced, both in the optical at 5875\AA\, and in the
near-IR at 1.083\,$\mu$m (which overlaps with P$\gamma$ but contributes most of the flux
seen at this date in the 1.08\,$\mu$m feature).
Note the broad N{\,\sc ii} feature blueward of He{\,\sc i}\,5875\AA, although its absorption
strength is somewhat overestimated (see Sect.~\ref{Sec_mod_disc}).


\subsubsection{Observation of the 3rd of November 1999}
\label{Sec_0311}

\begin{figure}[htp!]
\epsfig{file=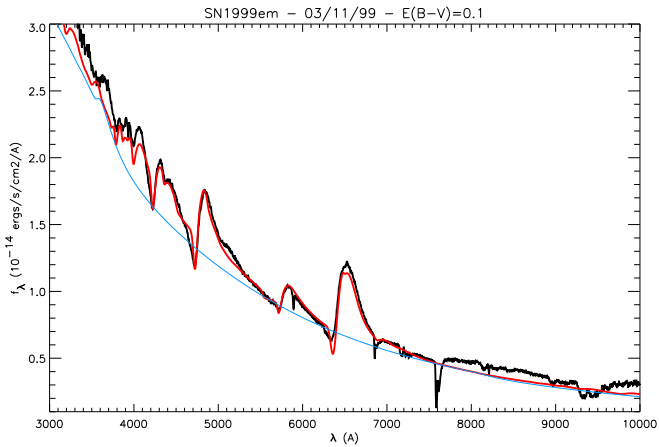, width=9cm}
\caption{
Synthetic fits (color) to observations (black) taken on November 3rd (H01; day 5).
The model has the following parameters:
$L_{\ast} = 9 \times 10^8 L_{\odot}$, $T_{\rm phot} = 10\,630$\,K,
$R_{\rm phot} = 6.7 \times 10^{14}$\,cm, $v_{\rm phot} = 8840$\,km\,s$^{-1}$,
$\rho_{\rm phot} = 3.6 \times 10^{-14}$\,g\,cm$^{-3}$, and $n=10$.
The model flux, which underestimates the observations by 50\%, 
is renormalized to the observed value at 5400\AA.
(This figure is available in color in the electronic version.)
}
\label{fig_0311}
\end{figure}

In Fig.~\ref{fig_0311}, we present a fit of the full (red) and continuum
(blue) synthetic SED to the observed SED of the 3rd of November 1999 (black; H01).
The model has the following properties:
$L_{\ast} = 9 \times 10^8 L_{\odot}$, $T_{\rm phot} = 10\,630$\,K,
$R_{\rm phot} = 6.7 \times 10^{14}$\,cm, $v_{\rm phot} = 8840$\,km\,s$^{-1}$,
$\rho_{\rm phot} = 3.6 \times 10^{-14}$\,g\,cm$^{-3}$, and $n = 10$.
The model flux, underestimated by 50\%, is renormalised to the observed value
at 5400\AA. The fit quality is good.
Increasing the outflow temperature (or ionization) results in a better
fit of the overall shape of the spectrum but leads to an overestimate of the
strength of He{\,\sc i}\,5875\AA\, and that of the N{\,\sc ii}\,5400\AA\, feature.
The bump in the red wing of H$\beta$, due primarily to Fe{\,\sc iii} lines,
is underestimated, suggesting perhaps a slightly higher environmental 
metallicity (Paper I).
As noted in the preamble to the section, we use, by default, a solar
mixture of metals.
Finally, note how reduced the continuum flux level is compared to the
full SED, highlighting the ubiquitous presence of broad and weak ``background'' lines.

\subsubsection{Observation of the 5th of November 1999}
\label{Sec_0511}

\begin{figure}[htp!]
\epsfig{file=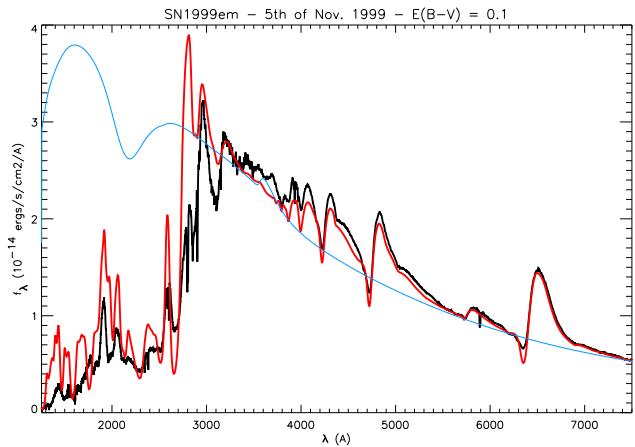, width=9cm}
\caption{
Full (red) and continuum (blue) synthetic fits to observations (black) taken 
on November 5th (UV range: Baron et al. 2000; optical range: L02, day 7).
The model has the following parameters:
$L_{\ast} = 5 \times 10^8 L_{\odot}$, $T_{\rm phot} = 9200$\,K, 
$R_{\rm phot} = 6.64 \times 10^{14}$\,cm, $v_{\rm phot} = 8750$\,km\,s$^{-1}$,
$\rho_{\rm phot} = 4.1 \times 10^{-14}$\,g\,cm$^{-3}$, and $n = 10$.
The model flux, which underestimates the observations by a factor of two, is renormalized 
to the observed value at 7500\AA. The dip in the continuum spectrum near
2200\,\AA\, is due to the effects of interstellar extinction.
(This figure is available in color in the electronic version.)
}
\label{fig_0511}
\end{figure}

In Fig.~\ref{fig_0511}, we show a full (red) and continuum (blue) synthetic
fit to observations taken on the 5th of November 1999 (L02), the optical range
being complemented with UV HST data (Baron et al. 2000; see Sect.~\ref{Sec_obs}).
The model has the following parameters:
$L_{\ast} = 5 \times 10^8 L_{\odot}$, $T_{\rm phot} = 9200$\,K, 
$R_{\rm phot} = 6.64 \times 10^{14}$\,cm, $v_{\rm phot} = 8750$\,km\,s$^{-1}$,
$\rho_{\rm phot} = 4.1 \times 10^{-14}$\,g\,cm$^{-3}$, and $n = 10$.
The model flux, which underestimates the observations by a factor of two, 
is renormalized to the observed value at 7500\AA.
Both observation and model were discussed in detail in Paper I (Sect. 3.2), 
emphasizing the complex and primary role played by
metal line-blanketing in the UV range (see lower panel of Fig. 3, Paper I).

The fit quality is good but there are noticeable discrepancies.
The optical flux is somewhat underestimated, while the UV flux is 
overestimated in a number of wavelength regions.
We overestimate the flux in the 2800\AA\, region, which, additionally to
Fe{\,\sc ii}, is strongly influenced by the multiplet of Mg{\,\sc ii},
composed of Mg{\,\sc ii}\,(3p-3s) and Mg{\,\sc ii}\,(4s-3p). 
Increasing/decreasing the Magnesium abundance by a 
factor of two introduces flux changes only at the 10\% level, thus
insufficient to resolve the discrepancy.
In Paper I, we showed how a global increase of the pristine metal 
abundance leads to a better match, e.g., of the strength of the 
iron feature at $\sim$5000\AA, giving some support for a solar or 
over-solar metallicity of the SN1999em environment; simultaneously, the increased 
metal-line blanketing in the UV reduces the emergent UV flux to more suitable levels.
Alternatively, lowering the effective temperature increases line-blocking 
through the shift from Fe{\,\sc iii} to Fe{\,\sc ii} opacity sources, also
matching the observed UV flux better; in this situation, however, the slope 
of the optical synthetic SED is then too flat.
We discuss further these issues in Sect.~\ref{Sec_mod_disc} but already here,
we see how coupled and subtle the effects of line-blanketing and outflow
temperature/ionization are.  
Note that the photospheric temperature of 9200\,K found here is in good agreement
with the value of 9000\,K proposed for that date by Baron et al. (2000).

Overall, having access to UV data during the early evolution is an asset, as it 
allows the determination, or the confirmation, of the reddening obtained usually
at other (relatively late) times from optical data.

\subsubsection{Observation of the 9th of November 1999}
\label{Sec_0911}

\begin{figure}[htp!]
\epsfig{file=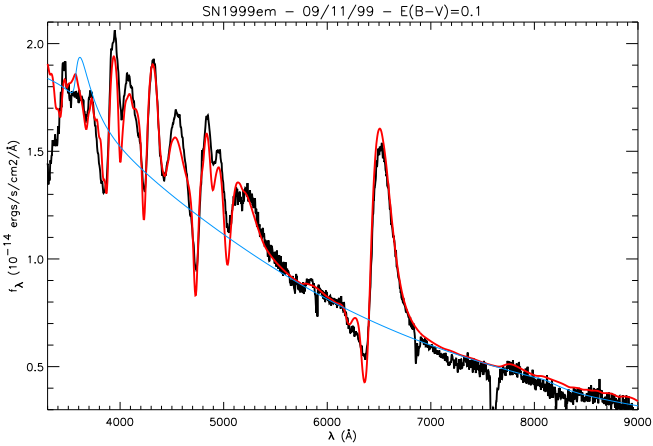, width=9cm}

\caption{
Synthetic fits (red) to observations (black) taken on November 9th (H01; day 11).
The model has the following parameters:
$L_{\ast} = 2.5 \times 10^8 L_{\odot}$, $T_{\rm phot} = 8040$\,K,
$R_{\rm phot} = 6.65 \times 10^{14}$\,cm, $v_{\rm phot} = 7960$\,km\,s$^{-1}$,
$\rho_{\rm phot} = 4 \times 10^{-14}$\,g\,cm$^{-3}$, and $n=10$.
The model flux, which underestimates the observations by a factor of 2.65, is renormalized to 
the observed value at 6000\AA. The feature in the continuum spectrum near 3600\,\AA\
arises from the inclusion, via level dissolution, of Balmer series bound-bound 
transitions near the series limit.
(This figure is available in color in the electronic version.)
}
\label{fig_0911}
\end{figure}

We show in Fig.~\ref{fig_0911} a synthetic fit to the observations of SN1999em
on the 9th of November 1999 (H01; day 11).
The model parameters are the following:
$L_{\ast} = 2.5 \times 10^8 L_{\odot}$, $T_{\rm phot} = 8040$\,K,
$R_{\rm phot} = 6.65 \times 10^{14}$\,cm, $v_{\rm phot} = 7960$\,km\,s$^{-1}$, 
$\rho_{\rm phot} = 4 \times 10^{-14}$\,g\,cm$^{-3}$, and $n=10$.
The model flux, which underestimates the observations by a factor of 2.65, 
is renormalized to the observed value at 6000\AA.
Starting around this date, metal line blanketing in the optical 
is noticeable, increasing steadily as we progress through the 
photospheric phase.
Mingled with the Balmer lines (H$\gamma$, H$\beta$, and H$\alpha$), one sees the 
presence of Ca{\,\sc ii} at 3800\AA\, (as for the previous date), Fe{\,\sc ii} 
and Ti{\,\sc ii} lines at $\sim$4400\AA, and Fe{\,\sc ii} lines redward 
of H$\beta$ up to 5500\AA. 
He{\,\sc i}\,5875\AA\, is still observable as a very weak feature,
soon to be replaced at the same location by the Na{\,\sc i}\,5890\AA\,
doublet.
Fe{\,\sc ii} and Si{\,\sc ii} are also present around 6300\AA, creating
a kink in the H$\alpha$ absorption trough; it might thus not be necessary 
to invoke a peculiar formation of the H$\alpha$ line to explain this 
feature (L02).

\subsubsection{Observation of the 14th of November 1999}
\label{Sec_1411}

\begin{figure}[htp!]
\epsfig{file=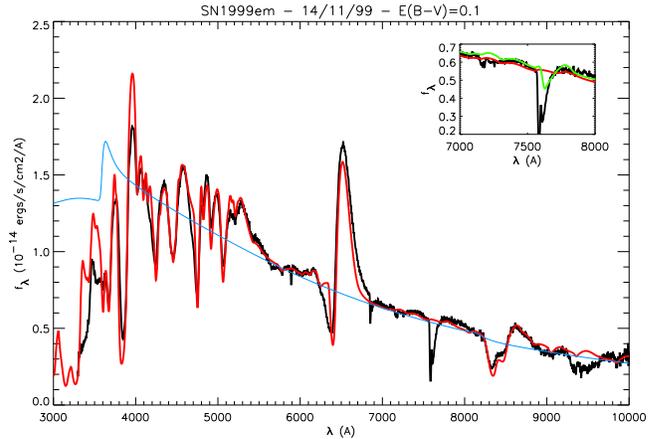, width=9cm}
\caption{
Full (red) and continuum (blue) synthetic spectral fits to observations (black) 
taken on November 14th (H01; day 16).
The model has the following parameters:
$L_{\ast} = 1.5 \times 10^8 L_{\odot}$, $T_{\rm phot} = 6800$\,K,
$R_{\rm phot}= 6.15 \times 10^{14}$\,cm, $v_{\rm phot} = 6350$\,km\,s$^{-1}$,
$\rho_{\rm phot} = 8.7 \times 10^{-14}$\,g\,cm$^{-3}$ and $n = 10$.
The model flux, which underestimates the observations by a factor of 3.7, is renormalized 
to the observed value at 6000\AA.
The species identification of line contributions is shown in the lower
panel of Fig.~5, Paper I. 
(This figure is available in color in the electronic version.)
}
\label{fig_1411}
\end{figure}

In Fig.~\ref{fig_1411}, we present full (red) and continuum (blue) synthetic
spectrum fits to observations taken on the 14th of November 1999 (H01; day 16).
The model parameters are the following:
$L_{\ast} = 1.5 \times 10^8 L_{\odot}$, $T_{\rm phot} = 6800$\,K,
$R_{\rm phot}= 6.15 \times 10^{14}$\,cm, $v_{\rm phot} = 6350$\,km\,s$^{-1}$,
$\rho_{\rm phot} = 8.7 \times 10^{-14}$\,g\,cm$^{-3}$, and $n = 10$.
The model flux, which underestimates the observations by a factor of 3.7, 
is renormalized to the observed value at 6000\AA.
Note that a detailed discussion of this model has been made in Sect.~3.3
of Paper I.

A conspicuous feature now appears in the $I$ band, attributable to
Ca{\,\sc ii}\,8498--8542--8662\AA, consistent with the
identification of Ca{\,\sc ii} 3933--3968\AA\ at earlier epochs.
He{\,\sc i}\,5875\AA\, has also vanished, replaced by a weak feature due to 
Na{\,\sc i}\,5890\AA, well fitted by invoking an enhancement
of four over the corresponding solar sodium abundance 
(Paper I; Prantzos et al. 1986).
The feature at 7700\AA\, now seems to extend too far to the red to be 
attributed exclusively to terrestrial atmospheric absorption, as supposed so far, and thus
suggests the presence of O{\,\sc i}\,7700\AA.
In the top right corner of Fig.~\ref{fig_1411}, we add, over the relevant 
region, the synthetic SED (green) obtained by enhancing the oxygen abundance 
by a factor of 100 over the standard value O/He = 10$^{-4}$, as
required to reproduce the red-side of the observed feature.
Observations at later epochs, revealing spectra with features both narrower and 
better centered on the rest-wavelength (see Paper I), will confirm this 
identification and the need for such a significant enhancement of the oxygen 
abundance over the CNO-cycle equilibrium value, characteristic of 
the deeper layers of the progenitor star (Hirshi et al. 2004).
Metal line features in the optical have also strengthened compared
to those of the previous date, following the cooling of the outflow, and
the recombination of hydrogen and Fe{\,\sc iii} (to Fe{\,\sc ii}).
The H$\alpha$ line profile is not too well fitted:
its red-wing extent is underestimated as is the P-Cygni trough, which is 
now very pronounced.
We instead predict a narrower profile shape, with considerable 
filling-in of the trough by Fe{\,\sc ii} and Si{\,\sc ii} lines.
As we will see below, this problem only gets worse, associated 
with the infamous underestimate of the H$\alpha$ strength.

\subsubsection{Observation of the 19th of November 1999}
\label{Sec_1911}

\begin{figure}[htp!]
\epsfig{file=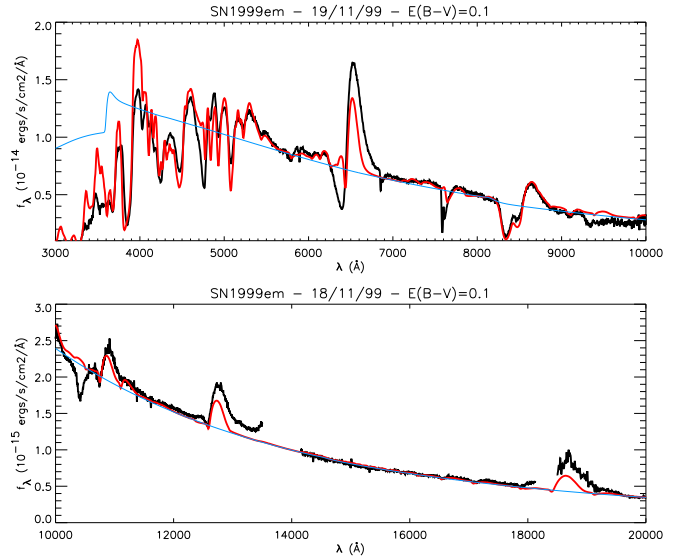, width=9cm}
%
\caption{
Synthetic fit (red) to observations (black) taken on November 19th (H01; optical
range, day 21) and 18th (H01; near-IR range).
The near-IR model flux is scaled down by 15\% compared to that in the optical,
to allow the simultaneous match to both wavelength regions.
The model properties are the following:
$L_{\ast} = 3.75 \times 10^8 L_{\odot}$, $T_{\rm phot} = 6260$\,K,
$R_{\rm phot} = 12.9 \times 10^{14}$\,cm, $v_{\rm phot} = 5530$\,km\,s$^{-1}$,
$\rho_{\rm phot} =  7.4 \times 10^{-14}$\,g\,cm$^{-3}$, and $n=10$.
The model flux, which underestimates the observations by 10\%, 
is renormalized to the observed value at 6100\AA.
(This figure is available in color in the electronic version.)
}
\label{fig_1911}
\end{figure}

We show in Fig.~\ref{fig_1911} synthetic fits (color) to the observations of
the 19th of November 1999 (H01; day 21).
The model flux is normalized to the observed one at 6100\AA\,
with an additional decrease of 15\% in the near-IR.
This flux shift is likely to result from the time difference
between the two observations rather than an inadequate
slope between the optical and the near-IR ranges in the
synthetic spectrum.
The model properties are the following:
$L_{\ast} = 3.75 \times 10^8 L_{\odot}$, $T_{\rm phot} = 6260$\,K,
$R_{\rm phot}= 12.9 \times 10^{14}$\,cm, $v_{\rm phot} = 5530$\,km\,s$^{-1}$,
$\rho_{\rm phot} =  7.4 \times 10^{-14}$\,g\,cm$^{-3}$, and $n = 10$.
The model flux, which underestimates the observations by 10\%, is renormalised to the observed 
value at 6100\AA.

As described in Paper I, we have found that using a small
turbulent velocity helps maintain significant emission
in H$\alpha$ (see next section).
In general, this entails using a turbulent velocity of 20\,km\,s$^{-1}$,
rather than 100\,km\,s$^{-1}$ used for models where hydrogen is fully or
mostly ionized.  Note however that this change only delays the onset of the
discrepancy for the H$\alpha$ fit, so is introduced here mostly for aesthetic reasons.
We find that lowering the density exponent down to 8 and even 6 leads to an
increase in the H$\alpha$ strength but also produces a very severe mismatch
of the Ca{\,\sc ii} lines, which become both stronger and broader than observed.
We find that indeed, maintaining a high density exponent of 10 leads to
better fits to observations, at least for the epochs covered here
and our model assumptions (Paper I).
Note also the presence of Fe{\,\sc ii} and Si{\,\sc ii} lines which fill in the 
H$\alpha$ trough.

Omitting this difficulty of reproducing H$\alpha$ and the Paschen series
the quality of the fit is again satisfactory, with a good match to the strength and shapes of 
line features due to Fe{\,\sc ii}, Ti{\,\sc ii}, Ca{\,\sc ii}.
The overall flux distribution is very well matched, with a slight
overestimate of the flux in the blue.
Another discrepancy is the slight overestimate of the Ca{\,\sc ii}/Fe{\,\sc ii}/Ti{\,\sc ii} 
blend around 3800\AA.  

We observe a weak feature at $\sim$1.07$\mu$m which we attribute to the 
C{\,\sc i}\,(3p-3s) multiplet.
However, to reproduce this weak feature, the carbon abundance had to be
increased by a factor of 10 to C/He = 0.0017, 
equivalent to a mass fraction of 0.0023.
Similarly, a number of C{\,\sc i} lines over the range 9060--9110\AA\,
are seen, although a firm identification is established with the observation of the 
5th of December 1999 (Sect.~\ref{Sec_0512}). 
Such a carbon abundance is compatible with chemical mixing prior to core-collapse,
inducing a more moderate depletion of C/O compared to N.
Accordingly, and following from the previous section, we identify the
O{\,\sc i}\,7700\AA\, line (adopting O/He = 10$^{-2}$),  as well as another oxygen 
feature in the near-IR at 1.13$\mu$m, standing in the red wing of P$\gamma$. 
Note that the latter also overlaps with a multiplets of Mg{\,\sc ii}\,(4p-3d) 
at 1.091-1.095$\mu$m (which represents about 30\% of the 1.09$\mu$m feature).

An observed feature at $\sim$1.03$\mu$m is not predicted by CMFGEN, despite
the numerous species included; this could come from singly-ionized species,
or perhaps Sc or Ba, not accounted for at present.

\begin{figure}[htp!]
\epsfig{file=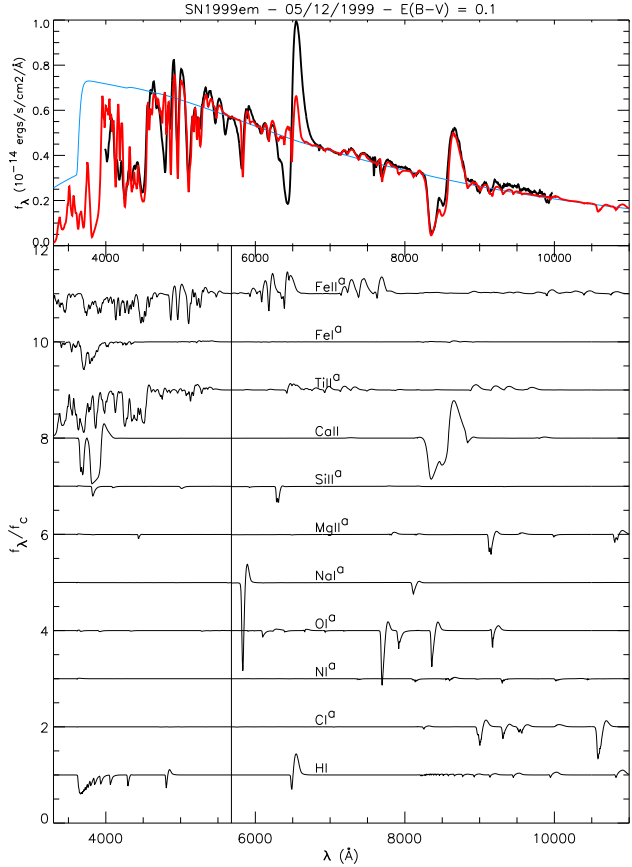, width=9cm}
%
\caption{
{\it Top}: full (red) and continuum (blue) synthetic spectrum fits to observations (black) taken
on the 5th of December 1999 (L02, day 37; this spectroscopic dataset is later used for distance determinations,
in combination with the photometry on day 38  for which no spectra exist.)
The model has the following parameters:
$L_{\ast} = 2 \times 10^8 L_{\odot}$, $T_{\rm phot} = 5920$\,K,
$R_{\rm phot} = 10.9 \times 10^{14}$\,cm, $v_{\rm phot} = 3400$\,km\,s$^{-1}$,
$\rho_{\rm phot} =  38.8 \times 10^{-14}$\,g\,cm$^{-3}$, and $n=10$.
The model flux, which underestimates the observations by 10\%, 
is renormalized to the observed value at 7000\AA.
{\it Bottom}:
montage of rectified spectra computed by including bound-bound transitions of 
individual ionization species, ordered from the bottom by increasing atomic weight.
For species labeled with an ``a'', flux departure from unity (continuum level)
is magnified by a factor of four beyond 5680\AA. 
(This figure is available in color in the electronic version.)
}
\label{fig_0512}
\end{figure}

\subsubsection{Observation of the 5th of December 1999}
\label{Sec_0512}

We present synthetic fits (color) to the observations (black) taken on the
5th of December 1999 (L02) in Fig.~\ref{fig_0512}.
The corresponding model has the following parameters:
$L_{\ast} = 2 \times 10^8 L_{\odot}$, $T_{\rm phot} = 5920$\,K,
$R_{\rm phot} = 10.9 \times 10^{14}$\,cm, $v_{\rm phot} = 3400$\,km\,s$^{-1}$,
$\rho_{\rm phot} =  38.8 \times 10^{-14}$\,g\,cm$^{-3}$, and $n=10$.
The model flux, which underestimates the observations by 10\%, 
is renormalized to the observed value at 7000\AA.
Also, to ease the identification of observed spectral features, made more numerous 
due to the reduced outflow expansion velocity compared to previous dates, 
we show, in the lower panel, rectified spectra obtained when only bound-bound
transitions of a given ionization species (see label) are included,
ordered upwards with increasing atomic mass.

The overall fit quality is good.
Discrepancies are noticeable for H$\alpha$, whose strength in absorption and emission
is underestimated; we are currently investigating possible extra energy sources
that could preserve the large spatial extent of Balmer/Paschen line formation, imposed by the 
morphology of H$\alpha$ at late times.
Contrary to previous epochs, the nitrogen abundance on this date has to be reduced to maintain
a good fit to the 8200\AA\, feature, whose strength is explained almost entirely by the
Na{\,\sc i} line (see below). We thus adopt N/He = 10$^{-3}$ (by number). 
The 7700\AA\, feature was at early-times associated with atmospheric
extinction alone, although on the 14th and 19th of November, the absorption seemed
to stretch further to the red, suggesting a contribution from the O{\,\sc i}(3p-3s) triplet at 7775\AA.
On the 5th of December 1999, following the reduced expansion velocity in the line formation
region, one can unambiguously observe a P-Cygni line profile shape, blending with the
atmospheric absorption only in the trough and blueward.  
As for earlier models, the reproduction of this line requires enhancing the oxygen abundance to 
O/He = 10$^{-2}$ (by number).
However, this leads to an overestimated strength for the O{\,\sc i}\,(3s-3p) triplet 
line at 7984\AA; other O{\,\sc i} lines at  8446\AA\, and 9264\AA\, overlap
with Ca{\,\sc ii}, or Mg{\,\sc ii} and C{\,\sc i} lines and are therefore 
not particularly useful as abundance constraints.
We retain the same carbon abundance as in Sect.~\ref{Sec_1911}, where the near-IR C{\,\sc i} feature
at 1.07$\mu$m required an enhancement, with C/He = 1.7 $\times$ 10$^{-3}$.
However, here, a clearly-defined C{\,\sc i} feature is observed, stemming from the (3p-3s) 
multiplet covering the range 9060--9110\AA.
We associate the 9200\AA\ feature with the Mg{\,\sc ii}\,(4p-4s) doublet at 9218--9244\AA, 
overlapping with the O{\,\sc i} multiplet at 9260\AA\, (note that we adopt a solar 
magnesium abundance).
Finally, the absorption strength of the Na{\,\sc i}\,(3p-3s) doublet at 5890--5895\AA\, is 
well matched with a factor of four enhancement compared to solar (Prantzos et al. 1986).

\subsection{Discussion of Model Results}
\label{Sec_mod_disc}

   The models presented in the previous section provide satisfactory fits to 
our sample of spectrophotometric observations of SN1999em. 
We now investigate the overall accuracy of these model parameters, focusing on 
chemical abundances (Sect.~\ref{Sec_chem_abund}), outflow ionization/temperature 
and reddening (Sect.~\ref{Sec_teff_red}), 
and photospheric velocity (Sect.~\ref{Sec_phot_vel}).

  \subsubsection{Chemical Abundances}
\label{Sec_chem_abund}

Using spectroscopic observations during the photospheric-phase evolution 
of type II SN, we anticipate inferring an outflow composition in conformity 
with CNO-cycle equilibrium values, typical of the envelope of the corresponding massive star 
progenitor, with at most traces of elements produced through explosive nucleosynthesis.
Indeed, adopting a blue- or red-supergiant surface composition 
(Prantzos et al. 1986) allows a good match to observed features of all species, 
for both their absolute and relative strengths.
But let us now investigate how such line strengths are affected by a relative modulation
of the outflow H/He/CNO chemistry, as would result from a distinct evolutionary stage
or chemical mixing efficiency (linked to, e.g., rotation and/or magnetic fields).

\begin{figure}[htp!]
\epsfig{file=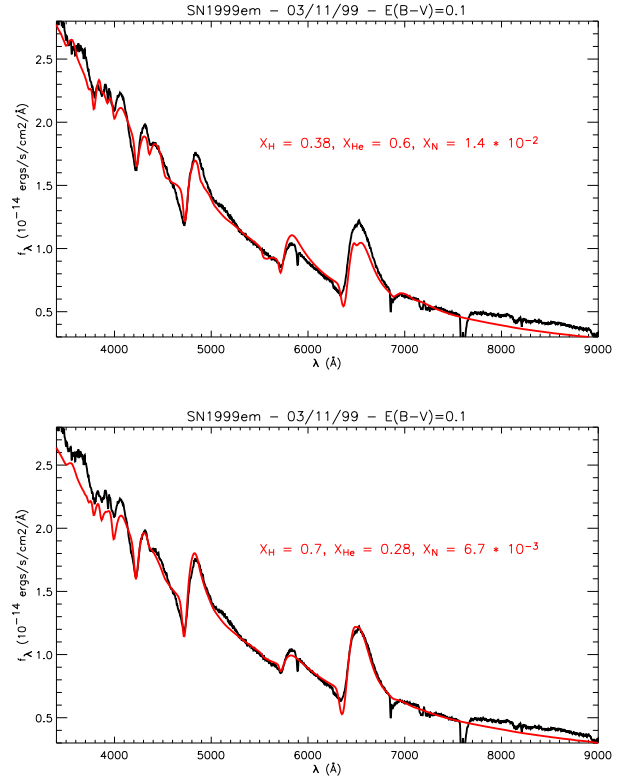, width=9cm}
%
\caption{
Synthetic fits (red) to observations (black) of SN1999em on 
the 3rd of November 1999 (H01; day 5) using a model with the same parameters 
as those given in Sect.~\ref{Sec_0311} (see also Fig.~\ref{fig_0311}) 
but having a more (top) or less (bottom) evolved chemical composition 
than the baseline model described in Sect.~\ref{Sec_mod_pres}.
(This figure is available in color in the electronic version.)
}
\label{fig_comp_hyd}
\end{figure}

We show in Fig.~\ref{fig_comp_hyd} a comparison between the observations of the 3rd of 
November 1999 (Sect.~\ref{Sec_0311}) and two synthetic 
spectra corresponding to more (top) or less (bottom) evolved progenitor stars.
For the former, we adopt  $X_{\rm H} = 0.38$, $X_{\rm He} = 0.60$, 
$X_{\rm C} = 3.1 \times 10^{-4}$, $X_{\rm N} = 1.4 \times 10^{-2}$, and 
$X_{\rm O} = 2.4 \times 10^{-4}$; for the latter, we use $X_{\rm H} = 0.7$,
$X_{\rm He} = 0.28$, $X_{\rm C} = 1.44 \times 10^{-4}$,
$X_{\rm N} = 6.7 \times 10^{-3}$, and $X_{\rm O} = 1.13 \times 10^{-4}$.

The modulation in hydrogen abundance translates into a variation of the density 
of free-electrons
in these fully ionized models: the higher the value of $X_{\rm H}$, the higher the electron-scattering 
optical depth, either at the photosphere or at the model base. Here, we find that the photosphere
for the $X_{\rm H} = 0.7$ model is located 5\% further out than the $X_{\rm H} = 0.38$ model,
at a cooler temperature (10\,270\,K rather than 11\,200\,K), explaining the slightly ``cooler''
SED for the hydrogen-richer model. 
Beside this slight change in the slope of the SED, the strength of individual features is 
noticeably altered.

In the  $X_{\rm H} = 0.38$ model, Balmer line strengths are significantly underestimated,
and He{\,\sc i}\,5875\AA\ overestimated. We also find that the
increased nitrogen abundance leads to the overestimate of the strength
of N{\,\sc ii} features around 5600\AA\, and 4600\AA. A comparison of Fig.~\ref{fig_comp_hyd}
with Fig.~\ref{fig_0311} shows that the depleted hydrogen model is clearly of inferior 
quality compared to our best fit model. With a solar H/He ratio the changes relative to our
best fit model are smaller, and it is more difficult to distinguish between the models.
The sensitivity to changes in hydrogen, helium, and nitrogen 
composition is to be contrasted, for example, with Eastman \& Kirshner (1989), who
required an unrealistically high helium abundance to reproduce the He{\,\sc i}\,5875\AA\,
line profile in the early spectra of SN1987A.

For carbon and oxygen, the situation is less clear: the outflow ionization is too low to
lead to C{\,\sc ii} or O{\,\sc ii} lines anywhere in the UV, optical, and near-infrared.
At a stage where hydrogen recombines in the outflow, the ionization is suitable for the
C{\,\sc i} line to appear around 9000\AA, whose fitting requires an enhancement of about a 
factor of ten above that of the baseline model (Sect.~\ref{Sec_0512}).
Similarly, the same carbon abundance enhancement is needed to explain the feature at 
1.07$\mu$m, observed in the spectrum of the 19th of November (Fig.~\ref{fig_1911}).
We find a similar situation for O{\,\sc i} lines, which appear in the red part of the
spectrum only at late times, and require a significant enhancement compared to
CNO-cycle equilibrium values.
Most notably, the O{\,\sc i} lines at 7700\AA\, and 1.13$\mu$m require to have
O/He in the vicinity of 0.01 (by number).

Overall, this investigation gives credence to our baseline model composition
(Sect.~\ref{Sec_mod_pres}) at early times, but with C/O abundances enhancing, 
as times progresses, above CNO-cycle equilibrium values; this suggests enhanced 
(rotational) mixing in the envelope of the progenitor star prior to collapse, 
or the appearance of the deeper and more chemically-evolved layers after just 
a few weeks past core-collapse.

  \subsubsection{Outflow Ionization, Effective Temperature, and Reddening}
\label{Sec_teff_red}

In the parameter space of photospheric-phase type II SN (Paper I), the similar 
effects produced by temperature and reddening modulations on the synthetic SED 
motivate a joint discussion.

\begin{figure}[htp!]
\epsfig{file=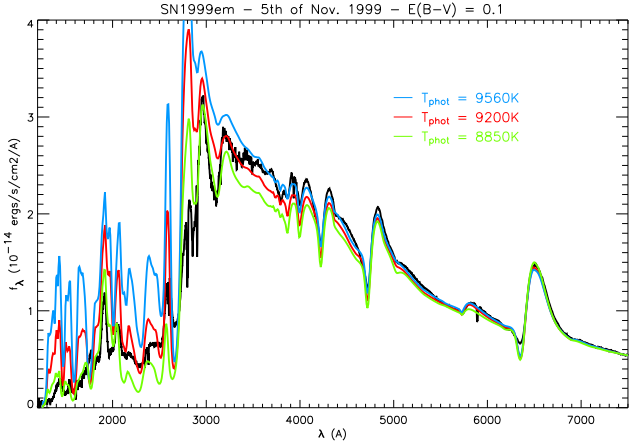, width=9cm}
\caption{
Illustration of the impact of a temperature change on the synthetic SED,
implemented through a 20\% increase (blue) or decrease (green) in 
luminosity (at constant base radius) compared to the model presented in 
Fig.~\ref{fig_0511} of Sect.\ref{Sec_0511} (red) and observations (black). 
All other model parameters are held fixed.
(This figure is available in color in the electronic version.)
}
\label{fig_comp_temp}
\end{figure}

In Fig.~\ref{fig_comp_temp}, we present spectroscopic observations of 
SN1999em taken on the 5th of November 1999 (Sect.~\ref{Sec_0511} and
Fig.~\ref{fig_0511}), but now overplotted with the SED from models 
with decreasing photospheric temperatures: $T_{\rm phot} = 9560$\,K (blue),
$T_{\rm phot} = 9200$\,K (red), and $T_{\rm phot} = 8850$\,K (green).
Within the corresponding range of outflow ionization, the reduction of the 
UV flux is not so much caused by the decrease in $T_{\rm phot}$, than 
by the opacity shift following the recombination from Fe{\,\sc iii} to Fe{\,\sc ii} 
in the SN outflow (as well as the opacity increase from numerous 
doubly-ionized metal species). 
In the optical, this shift has little bearing and the synthetic SED
changes solely through the modest variation in $T_{\rm phot}$.

\begin{figure}[htp!]
\epsfig{file=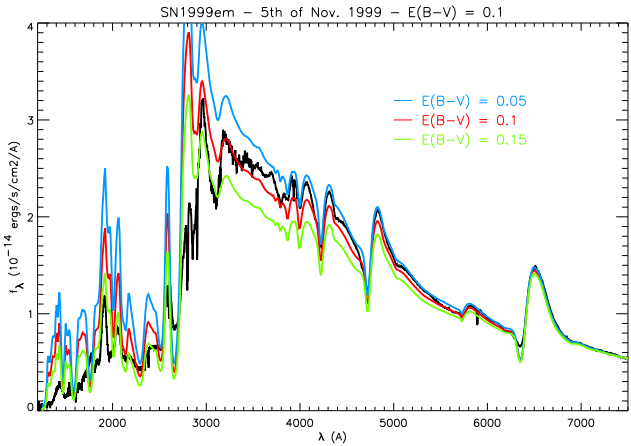, width=9cm}
\caption{
Illustration of the effect of reddening modulations on the SED.
We show the model and observations for the 5th of November 1999
(Fig.~\ref{fig_0511}, day 7), reddening the synthetic SED with $E(B-V)$ of
0.05 (blue), 0.1 (red), and 0.15 (green).
Note how, in this parameter range, a 0.05 change in $E(B-V)$ is comparable 
to a $\sim$$\pm$\,500\,K shift in photospheric temperature (Fig.~\ref{fig_comp_temp}).
(This figure is available in color in the electronic version.)
}
\label{fig_comp_red}
\end{figure}

Note, however, that applying a reddening modulation to the model corresponding
to the red line in Fig.~\ref{fig_comp_red} leads to a similar modulation:
the relatively-enhanced UV-extinction (Cardelli et al. 1988) echoes the strong 
line-blanketing due to metals discussed above.
The relatively-reduced optical-extinction echoes the weak SED sensitivity to 
small $T_{\rm phot}$ variations.
Overall, a change in $E(B-V)$ of 0.05 or in $T_{\rm phot}$ of 
$\sim$$\pm$\,500\,K have comparable impacts on the SED in the present parameter 
space. 

Early-epoch multi-wavelength observations are thus very important:
metal line-blanketing is then only moderate and helps disentangle
between temperature and reddening effects; The UV range is also essential
since this is where tests on the extinction magnitude are most sensitive.

  \subsubsection{Photospheric velocity}
  \label{Sec_phot_vel}

In Sect. 5 of Paper 2, we discussed how the velocity at maximum absorption
in a sample of optical-range synthetic P-Cygni line profiles (i.e. Balmer lines
and Fe\two5169\AA) relates to the corresponding model photospheric velocity.
Contrary to the usually held belief, such a measurement on optically-thick
lines does not systematically overestimate the model photospheric velocity.
For any given ray, with impact parameter $p$, the location of maximum
absorption for the flux-like quantity $p \cdot I(p)$ is exterior to the
(continuum) photosphere, but for increasing $p$, this location shifts to smaller
line-of-sight velocities, ultimately reaching line center.
The total line flux is the integral of $p \cdot I(p)$ over all $p$, and thus
reflects this range of velocities for the region of maximum-absorption.
We find that although the velocity at maximum absorption for an
optically-thick line always overestimates $v_{\rm phot}$ {\it for the ray
with impact parameter p=0}, it can either overestimate or underestimate
$v_{\rm phot}$ when the total line profile is considered.
In practice, we identified few parameters that intervene in this correlation,
the most important being the optical-depth of the line considered, which
is related to the outflow ionization state, and the density gradient in the
line formation region.
These are physical entities that are difficult to constrain without
detailed modeling.

Second, P-Cygni line profiles have shapes that make the location of maximum
absorption difficult to identify.
This problem is severe at early times (say for the first week after
explosion), when the expansion velocity in the continuum- and line-forming regions 
is much larger and the flux in the lines
departs little from that in the continuum, making line features quite weak.
This can be understood in the following way: suppose we take a model and
increase its photospheric velocity: to first order, this will not change
level populations, outflow ionization etc... and thus the total flux in
all lines will be essentially the same, but the flux per-velocity-bin 
will be reduced, the more so for larger velocities.
In practice, inspection of the first spectra for, e.g., SN1987A and SN1999em,
show both very broad and very weak P-Cygni troughs, the region of
maximum absorption covering few thousand km\,s$^{-1}$.
In view of the argument discussed in the preceding paragraph, it is then
difficult to infer a meaningful and accurate velocity for the photosphere.

Third, the assumption of a unique radius for the photosphere might
not be appropriate on physical grounds.
Multi-dimensional radiation-hydrodynamics simulations of core-collapse SN
(Burrows et al. 1995; Janka \& M\"{u}ller 1996) predict explosions with a systematic departure
from sphericity, both from the large-scale (following, e.g., the core and envelope
rotation) and small-scale (following, e.g., Rayleigh-Taylor instabilities) viewpoints.
Such effects are expected to make the photosphere quite ``fuzzy" at later times (say one month
after explosion), showing a range of spatial scales for different lines of
sight and a significant dispersion in the expansion velocity.

\begin{figure}[htp!]
\epsfig{file=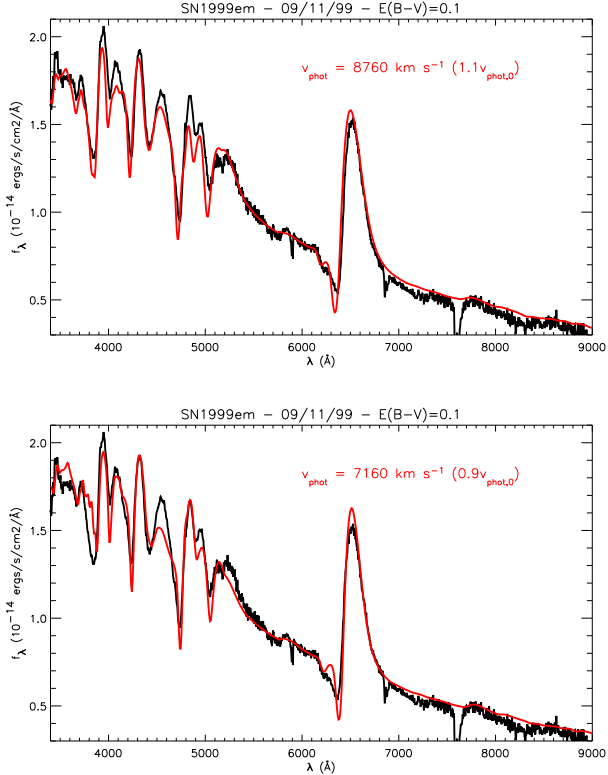, width=9cm}
\caption{
{\it Top}: synthetic fits (red) to observations (black) of SN1999em on the 9th of November 1999
(H01; day 11) using a model with the same parameters as those given in
Sect.~\ref{Sec_0911} (see also Fig.~\ref{fig_0911}) except for a photospheric velocity
enhanced by 10\% ($v_{\rm phot} = 8760$\,km\,s$^{-1}$, $v_{\rm phot,0} = 7960$\,km\,s$^{-1}$).
{\it Bottom}: same as above, but this time the photospheric velocity is reduced by 10\%
($v_{\rm phot} = 7160$\,km\,s$^{-1}$) compared to that of the reference model of Sect.~\ref{Sec_0911}.
(This figure is available in color in the electronic version.)
}
\label{fig_comp_vel}
\end{figure}

To illustrate the difficulty in assessing the photospheric velocity
from P-Cygni line profiles, we show in Fig.~\ref{fig_comp_vel} synthetic
fits (red) to the observation (black) of SN1999em taken on the 9th of November 1999.
The models shown have identical parameters to the one presented in
Sect.~\ref{Sec_0911} departing only from its photospheric velocity $v_{\rm
phot, 0}=7960$\,km\,s$^{-1}$: if the model photospheric velocity
is 10\% higher (top), the H$\alpha$ width is overestimated but H$\beta$,
Ca\two\,3800\AA, and Fe\two\,5159\AA\, are better fitted than in the 
model with a 10\% slower photospheric velocity (bottom). 
But all three models fit almost equally well the observations so that there is
a physically-meaningful 
uncertainty in photospheric velocity at the 10\% level. While it is possible to
measure precise velocities from the location of maximum absorption in some P~Cygni profiles, 
the errors associated with such measurements significantly underestimate the errors in the inferred
photospheric velocity.

Thus, the inference of the photospheric velocity is made difficult 
by the ambiguity of interpreting line width and velocity at maximum absorption in P-Cygni
profiles, as well as, perhaps, the physical velocity range spanned by the line formation 
region of different species.


For later use in distance determinations, we thus choose the
model photospheric velocity obtained in Sect.~\ref{Sec_mod_analysis}
and given in Table~\ref{tab_mod_analysis};
direct measurements on the observed spectra are more precise but the correspondence
with $v_{\rm phot}$ is unclear.
For all dates, the overall quality of our fits is good, which gives us confidence that our model
photospheric velocity is a meaningful reflection of the expansion velocity of the photosphere, 
say to within $\pm$10\% at most.

\section{The Distance to SN1999em}
\label{Sec_dist}

Let us now determine the distance $D$ to SN1999em --- to do so we will use
5 distinct, but closely related, approaches. For the first 2 approaches
we use the ``standard'' form of the EPM (H01, L02), in combination with either 
E96/H01's or Paper II's correction factors. In the third approach we use the models
to derive the temperatures which are then combined with Papers II's correction
factors. With the final EPM approach, we use both temperatures and
correction factors inferred from the best fit models.
Finally, we use the SEAM approach of Baron et al. (2004).


As discussed in Sects.~\ref{Sec_obs}-\ref{Sec_mod}, we selected eight epochs 
(October 30th, November 1st, 3rd, 5th, 9th, 14th, 19th, and December 5th)
for which CMFGEN fitted well both continuum and line fluxes.
Relative to the discovery date HJD 2,451,480.94, these correspond to 
days 1, 3, 5, 7, 11, 16, 21, and 38.
No spectra were taken on day 38: for that date, given the very slow spectroscopic and photometric 
evolution at such late times, we combine the spectrum on day 37 with the photometry of day 38.
Whatever the approach, we use the relevant model photospheric velocity.
Note also that, in this section, all synthetic magnitudes are computed 
using the transmission functions and zero points of H01.

\subsection{Inference with the EPM}
\label{Sec_epm}

A key component of the EPM (E96, H01, L02, Paper II) lies in the correspondence 
between correction factors and blackbody color temperatures $T_{\rm c}$,
defined from detailed models of type II SN spectra (E96; Paper II).
In Paper II, we showed that our correction factors possess similar properties to 
those found in E96, albeit offset upwards by $\sim$20\%.
At a given $T_{\rm c}$, correction factors show an intrinsic scatter
of $\sim$10-20\%, roughly independent of $T_{\rm c}$. 
However, if caused by uncertainties in $T_{\rm c}$, this scatter increases dramatically at low-$T_{\rm c}$.
Tabulated correction factors may also be inadequate to account for the blackbody 
assumption, if changes in magnitudes are due to variations in expansion velocity
or metallicity, rather than changes, e.g., in temperature or the strength of flux dilution.
Hence, we anticipate that using tailored models for each individual
date will be more consistent and thereby improve the use of the EPM 
technique for distance determinations.

In this section, we show a sequence of computations of distances based on distinct
sets of quantities $\theta_S$, $\xi_S$ and $T_S$.
Here, $S$ refers to one of the three bandpass combinations, $\{B,V\}$, $\{B,V,I\}$, 
and $\{V,I\}$, $\theta_S$ to the quantity $R_{\rm phot}/D$ (not known individually),   
$\xi_S$ to correction factors, and $T_S$ to the blackbody color temperature, each 
of the last three quantities corresponding to a selected bandpass combination $S$. 
An error analysis of the EPM technique and a description of the technique
used to derive the distances are given in Appendix A \& B.

In Table~\ref{tab_epm1}, we show such a set computed with
the ``standard'' form of the EPM (L02, H01) and using H01's correction factors.
Restricting first the observational dataset to the first seven dates (prior
to and excluding the 5th of December 1999)\footnote{
Throughout the paper we use observational data sets containing either 7 or
8 dates. This serves two purposes: First, H$\alpha$ in the last data set
is not well fitted, meaning that systematic errors could, potentially, be
large. Second, the effect of the longer time baseline, which
helps to increase the accuracy of the explosion date, can be discerned.},
we obtain increasing distances
for the three bandpass combinations, $\{B,V\}$, $\{B,V,I\}$, and $\{V,I\}$,
from 8.6\,Mpc, to 9.7\,Mpc, and 11.7\,Mpc, while the explosion date $t_{\rm exp}$ 
moves back in time in the same order, from -4.4 days, to -7.0 days, and -10.4 days.
The disparity amongst bandpass combinations is large, especially for the 
explosion dates, and unsatisfactory since the EPM should yield similar distances 
and explosion times whatever choice of bandpass combination, 
provided correction factors across the range of 
color temperatures are appropriate.
Adding the eighth date to the dataset leads to systematically
smaller distances which are in better agreement but the explosion dates 
still differ substantially. The derived distance is approximately 8.7 Mpc which
is 75\% of the Cepheid-distance to NGC\,1637 and SN1999em. 


\begin{table*}
\caption[]{Table of EPM quantities $\theta_S$ (10$^8$\,km\,Mpc$^{-1}$), $\xi_S$, and
$T_S$ (K), for the eight epochs covered by our analysis (i.e. from the 30th of October till
the 5th of December 1999; see Sect.~\ref{Sec_obs}). Also given are the resulting distance
and explosion dates inferred. Days and dates are relative to JD 2,451,480.94}
\label{tab_epm1}
\begin{center}
\begin{tabular}{cccccccccccccccccccccc}
\hline
\multicolumn{10}{c}{EPM combined with Minimization Technique
and $\xi$ prescription from H01} \\
\hline
Day & \multicolumn{3}{c}{Angular size}
& \multicolumn{3}{c}{Correction Factor}
& \multicolumn{3}{c}{Temperature} \\
& \multicolumn{3}{c}{(10$^8$\,km\,Mpc$^{-1}$)}
& \multicolumn{3}{c}{}
& \multicolumn{3}{c}{(K)} \\
\hline
&
$\theta_{BV}$ & $\theta_{BVI}$ & $\theta_{VI}$ &
$\xi_{BV}$    & $\xi_{BVI}$    & $\xi_{VI}$    &
$T_{BV}$      & $T_{BVI}$      & $T_{VI}$      & \\
\hline
  1.0 &  6.28 &  7.78 &  8.84 &   0.405 & 0.425 & 0.440 &  16860& 13513& 11543 \\
  3.0 &  6.73 &  8.31 &  9.37 &   0.401 & 0.423 & 0.440 &  16476& 13241& 11383 \\
  5.0 &  8.16 &  9.14 &  9.82 &   0.382 & 0.416 & 0.438 &  14394& 12312& 10998 \\
  7.0 &  8.46 &  9.89 & 10.87 &   0.377 & 0.413 & 0.436 &  13801& 11479& 10070 \\
 11.0 & 11.10 & 11.48 & 11.85 &   0.367 & 0.412 & 0.436 &  11271& 10214&  9477 \\
 16.0 & 15.77 & 13.89 & 12.23 &   0.461 & 0.434 & 0.437 &   7747&  8532&  9365 \\
 21.0 & 16.45 & 15.47 & 13.89 &   0.600 & 0.478 & 0.445 &   6514&  7411&  8404 \\
 38.0 & 17.51 & 17.58 & 16.98 &   1.202 & 0.652 & 0.482 &   4688&  5777&  7043 \\
\hline
\hline
      & \multicolumn{3}{c}{$BV$ Set}& \multicolumn{3}{c}{$BVI$ Set} & \multicolumn{3}{c}{$VI$ Set} \\
\hline
 & \multicolumn{9}{c}{Using the first 7 dates only} \\
\hline
  $D$ &  \multicolumn{3}{c}{$8.6 \pm 0.8$}&  \multicolumn{3}{c}{$9.7 \pm 1.0$} &  \multicolumn{3}{c}{$11.7 \pm 1.5$}  \\
$t_{\rm exp}$ &  \multicolumn{3}{c}{$-4.4\pm 0.9$}&  \multicolumn{3}{c}{$-7.0 \pm 1.4$} &  \multicolumn{3}{c}{$-10.4 \pm 2.1$}  \\
\hline
 & \multicolumn{9}{c}{Using 8 dates} \\
\hline
  $D$ &  \multicolumn{3}{c}{$8.0 \pm 0.6$}&  \multicolumn{3}{c}{$8.8 \pm 0.7$} &  \multicolumn{3}{c}{$10.1 \pm 0.9$ }  \\
$t_{\rm exp}$ &  \multicolumn{3}{c}{$-3.9 \pm 0.7$}&  \multicolumn{3}{c}{$-5.9 \pm 1.0$} &  \multicolumn{3}{c}{$-8.4 \pm 1.4$}  \\
\hline
\end{tabular}
\end{center}
\end{table*}

\begin{table*}
\caption[]{Same as Table~\ref{tab_epm1}, this time using Paper II's $\xi$-prescription}
\label{tab_epm2}
\begin{center}
\begin{tabular}{cccccccccccccccccccccc}
\hline
\multicolumn{10}{c}{EPM combined with Minimization Technique
and $\xi$ prescription from Paper II} \\
\hline
Day & \multicolumn{3}{c}{Angular size}
& \multicolumn{3}{c}{Correction Factor}
& \multicolumn{3}{c}{Temperature} \\
& \multicolumn{3}{c}{(10$^8$\,km\,Mpc$^{-1}$)}
& \multicolumn{3}{c}{}
& \multicolumn{3}{c}{(K)} \\
\hline
&
$\theta_{BV}$ & $\theta_{BVI}$ & $\theta_{VI}$ &
$\xi_{BV}$    & $\xi_{BVI}$    & $\xi_{VI}$    &
$T_{BV}$      & $T_{BVI}$      & $T_{VI}$      & \\
\hline
  1.0 &  5.82 &  6.58 &   7.41 &   0.436 & 0.504&  0.526 &  16876& 13481& 11543 \\
  3.0 &  6.20 &  6.95 &   7.86 &   0.439 & 0.505&  0.525 &  16364& 13241& 11351 \\
  5.0 &  6.88 &  7.48 &   8.16 &   0.453 & 0.508&  0.525 &  14378& 12328& 11031 \\
  7.0 &  6.95 &  7.93 &   8.99 &   0.460 & 0.514&  0.526 &  13769& 11495& 10086 \\
 11.0 &  8.09 &  8.99 &   9.74 &   0.503 & 0.530&  0.530 &  11271& 10166&  9493 \\
 16.0 & 10.57 & 10.50 &  10.12 &   0.688 & 0.572&  0.532 &   7747&  8548&  9317 \\
 21.0 & 11.63 & 11.70 &  11.25 &   0.854 & 0.632&  0.547 &   6498&  7411&  8420 \\
 38.0 & 14.87 & 13.96 &  13.51 &   1.414 & 0.820&  0.606 &   4688&  5777&  7043 \\
\hline
\hline
      & \multicolumn{3}{c}{$BV$ Set}& \multicolumn{3}{c}{$BVI$ Set} & \multicolumn{3}{c}{$VI$ Set} \\
\hline
 &\multicolumn{9}{c}{Using the first 7 dates only} \\
\hline
  $D$ &  \multicolumn{3}{c}{$13.5 \pm 1.5$}&  \multicolumn{3}{c}{$12.5 \pm 1.6$} &  \multicolumn{3}{c}{$14.6 \pm 1.9$}  \\
$t_{\rm exp}$ &  \multicolumn{3}{c}{$-7.2\pm 1.4$}&  \multicolumn{3}{c}{$-8.4\pm1.7$} &  \multicolumn{3}{c}{$-10.1 \pm 2.2$}  \\
\hline
 & \multicolumn{9}{c}{Using 8 dates} \\
\hline
  $D$ &  \multicolumn{3}{c}{$11.7 \pm 1.0$}&  \multicolumn{3}{c}{$11.9 \pm 1.0$} &  \multicolumn{3}{c}{$12.6 \pm 1.2$}  \\
$t_{\rm exp}$ &  \multicolumn{3}{c}{$-5.7\pm 1.0$}&  \multicolumn{3}{c}{$-6.9\pm1.2$} &  \multicolumn{3}{c}{$-8.8 \pm 1.5$}  \\
\hline
\end{tabular}
\end{center}
\end{table*}

\begin{table*}
\caption[]{Same as Table~\ref{tab_epm1}, this time using blackbody color temperatures
from Sect.~\ref{Sec_mod}, and Paper II's $\xi$-prescription}
\label{tab_epm3}
\begin{center}
\begin{tabular}{cccccccccccccccccccccc}
\hline
\multicolumn{10}{c}{EPM combined with Minimization Technique,} \\
\multicolumn{10}{c}{blackbody color temperatures from Sect.~\ref{Sec_mod},}\\
\multicolumn{10}{c}{and $\xi$-prescription from Paper II} \\
\hline
Day & \multicolumn{3}{c}{Angular size}
& \multicolumn{3}{c}{Correction Factor}
& \multicolumn{3}{c}{Temperature} \\
& \multicolumn{3}{c}{(10$^8$\,km\,Mpc$^{-1}$)}
& \multicolumn{3}{c}{}
& \multicolumn{3}{c}{(K)} \\
\hline
&
$\theta_{BV}$ & $\theta_{BVI}$ & $\theta_{VI}$ &
$\xi_{BV}$    & $\xi_{BVI}$    & $\xi_{VI}$    &
$T_{BV}$      & $T_{BVI}$      & $T_{VI}$      & \\
\hline
  1.0  &   6.13  &   5.90  &   5.67  &  0.440  &  0.503 &   0.543 & 16024 &  14965 &  14388 \\
  3.0  &   6.35  &   6.05  &   5.97  &  0.441  &  0.503 &   0.542 & 15959 &  15030 &  14228 \\
  5.0  &   6.95  &   6.95  &   7.03  &  0.455  &  0.505 &   0.530 & 14228 &  13202 &  12529 \\
  7.0  &   7.33  &   7.63  &   7.86  &  0.470  &  0.510 &   0.525 & 12945 &  11951 &  11278 \\
 11.0  &   8.46  &   8.91  &   9.37  &  0.519  &  0.529 &   0.527 & 10701 &  10252 &   9835 \\
 16.0  &  10.12  &  10.12  &   9.82  &  0.650  &  0.562 &   0.529 &  8168 &   8841 &   9547 \\
 21.0  &  10.72  &  11.03  &  10.80  &  0.760  &  0.605 &   0.540 &  7110 &   7847 &   8745 \\
 38.0  &  12.08  &  12.76  &  13.21  &  1.077  &  0.734 &   0.594 &  5539 &   6340 &   7238 \\
\hline
\hline
      & \multicolumn{3}{c}{$BV$ Set}& \multicolumn{3}{c}{$BVI$ Set} & \multicolumn{3}{c}{$VI$ Set} \\
\hline
 & \multicolumn{9}{c}{Using the first 7 dates only} \\
\hline
  $D$ &  \multicolumn{3}{c}{$14.2 \pm 1.6$} &  \multicolumn{3}{c}{$13.2\pm 1.4$} &  \multicolumn{3}{c}{$12.9\pm1.3$}  \\
$t_{\rm exp}$ &  \multicolumn{3}{c}{$-8.2\pm 1.6$}&  \multicolumn{3}{c}{$-7.2 \pm 1.4$} &  \multicolumn{3}{c}{$-6.9 \pm 1.3$}  \\
\hline
 & \multicolumn{9}{c}{Using 8 dates} \\
\hline
  $D$ &  \multicolumn{3}{c}{$12.8 \pm 1.1$} &  \multicolumn{3}{c}{$12.0 \pm 1.0$} &  \multicolumn{3}{c}{$11.7\pm 0.9$}  \\
$t_{\rm exp}$ &  \multicolumn{3}{c}{$-8.0\pm 1.2$}&  \multicolumn{3}{c}{$-6.2 \pm 1.$} &  \multicolumn{3}{c}{$-5.9 \pm 1.0$}  \\
\hline
\end{tabular}
\end{center}
\end{table*}

\begin{table*}
\caption[]{Same as Table~\ref{tab_epm1}, this time using blackbody color temperatures
and $\xi$-values from Sect.~\ref{Sec_mod}}
\label{tab_epm4}
\begin{center}
\begin{tabular}{cccccccccccccccccccccc}
\hline
\multicolumn{10}{c}{EPM combined with blackbody color temperatures} \\
\multicolumn{10}{c}{and $\xi$-values from Sect.~\ref{Sec_mod}} \\
\hline
Date & \multicolumn{3}{c}{Angular size}
& \multicolumn{3}{c}{Correction Factor}
& \multicolumn{3}{c}{Temperature} \\
& \multicolumn{3}{c}{(10$^8$\,km\,Mpc$^{-1}$)}
& \multicolumn{3}{c}{}
& \multicolumn{3}{c}{(K)} \\
\hline
&
$\theta_{BV}$ & $\theta_{BVI}$ & $\theta_{VI}$ &
$\xi_{BV}$    & $\xi_{BVI}$    & $\xi_{VI}$    &
$T_{BV}$      & $T_{BVI}$      & $T_{VI}$      & \\
\hline
  1.0  &   5.82 &    5.90 &    5.97  &    0.462 &   0.500  &  0.519  &   16024 &  14965 &  14388 \\
  3.0  &   6.50 &    6.58 &    6.65  &    0.430 &   0.462  &  0.488  &   15959 &  15030 &  14228 \\
  5.0  &   7.11 &    7.18 &    7.26  &    0.443 &   0.488  &  0.513  &   14228 &  13202 &  12529 \\
  7.0  &   7.78 &    7.86 &    7.86  &    0.443 &   0.494  &  0.526  &   12945 &  11951 &  11278 \\
 11.0  &   9.37 &    9.44 &    9.37  &    0.469 &   0.500  &  0.526  &   10701 &  10252 &   9835 \\
 16.0  &  10.20 &   10.20 &   10.27  &    0.646 &   0.558  &  0.507  &    8168 &   8841 &   9547 \\
 21.0  &  11.85 &   12.01 &   12.16  &    0.685 &   0.558  &  0.481  &    7110 &   7847 &   8745 \\
 38.0  &  12.16 &   12.38 &   12.83  &    1.072 &   0.755  &  0.608  &    5539 &   6340 &   7238 \\
\hline
\hline
      & \multicolumn{3}{c}{$BV$ Set}& \multicolumn{3}{c}{$BVI$ Set} & \multicolumn{3}{c}{$VI$ Set} \\
\hline
 & \multicolumn{9}{c}{Using the first 7 dates only} \\
\hline
  $D$ &  \multicolumn{3}{c}{$12.4\pm 1.2$} &  \multicolumn{3}{c}{$12.4 \pm 1.3$} &  \multicolumn{3}{c}{$12.4 \pm 1.3$}  \\
$t_{\rm exp}$ &  \multicolumn{3}{c}{$-6.8\pm 1.3$}&  \multicolumn{3}{c}{$-6.9\pm 1.$} &  \multicolumn{3}{c}{$-7.0\pm 1.3$}  \\
\hline
 & \multicolumn{9}{c}{Using 8 dates} \\
\hline
  $D$ &  \multicolumn{3}{c}{$11.7 \pm 1.0$} &  \multicolumn{3}{c}{$11.6 \pm 1.0$} &  \multicolumn{3}{c}{$11.5 \pm 0.9$}  \\
$t_{\rm exp}$ &  \multicolumn{3}{c}{$-6.2\pm 1.0$}&  \multicolumn{3}{c}{$-6.2\pm 1.0$} &  \multicolumn{3}{c}{$-6.2 \pm 1.0$}  \\
\hline
\end{tabular}
\end{center}
\end{table*}

We show in Table~\ref{tab_epm2} the corresponding quantities obtained with
Paper II's $\xi$-prescription.
These new correction factors are indeed always higher than those shown
in Table~\ref{tab_epm1} for the same date, but the magnitude of the difference varies
from date to date.
Hence, despite the preserved blackbody color temperatures between
Table~\ref{tab_epm1}-\ref{tab_epm2}, the non-uniform offset in correction factors between
the two prescriptions induces a non-uniform offset in individual $\theta_S$.
Resulting distances are systematically larger, using either 7 or 8 dates,
and the dispersion in the explosion times is reduced.

We have investigated whether enforcing a {\it fixed} scaling $x > 1$ on correction factors
given by H01 would clarify these behaviors.
As above, we find it introduces no noticeable 
modulation of the inferred color temperature (these
remain constant at the one percent level).
However, the individual values $\theta_S$ obtained are then scaled by $1/x$, basically leaving
$\xi_S \times \theta_S$ constant.
The resulting distances in that case are also scaled by $x$, but the explosion dates
are now essentially unchanged.
Using H01 or Paper II correction factors thus modifies {\it both} the inferred
time of explosion and the distance, an effect arising from the {\it non-uniform} scaling in
correction factors between H01 and Paper II; in other words, the differential variation of $\xi_S$
with $T_S$ affects distinct dates differentially.

An important factor influencing the accuracy of the distance determinations
is the incompatibility, at a given date, between color
temperatures $T_S$ obtained here (Table~\ref{tab_epm1}-\ref{tab_epm2}),
and those previously computed on corresponding CMFGEN model SEDs (Table~\ref{tab_mod_analysis} in
Sect.~\ref{Sec_mod_analysis}).
For bandpass combinations $\{B,V,I\}$ and $\{V,I\}$, model values of $T_S$ are
systematically larger 
than those deduced from the minimization technique performed on observations, while
for $\{B,V\}$, they are smaller at high and larger at low $T_{BV}$. 
This color-mismatch is likely to stem from the flux offset between synthetic
and observed spectra, leading to an over- or under-estimate of the corresponding 
bandpass magnitudes on which the computation of color-temperatures is based.
Figure~\ref{fig_0311} shows such a situation where the observed $I$ band flux 
is substantially underestimated by our model - note that the flux calibration
may also be at fault, especially at longer optical wavelengths where atmospheric extinction
operates.
To be more quantitative, we provide an estimate of the effect of flux errors on the 
derived angular diameter in Appendix A (Table~A.1).


\begin{figure}[htp!]
\epsfig{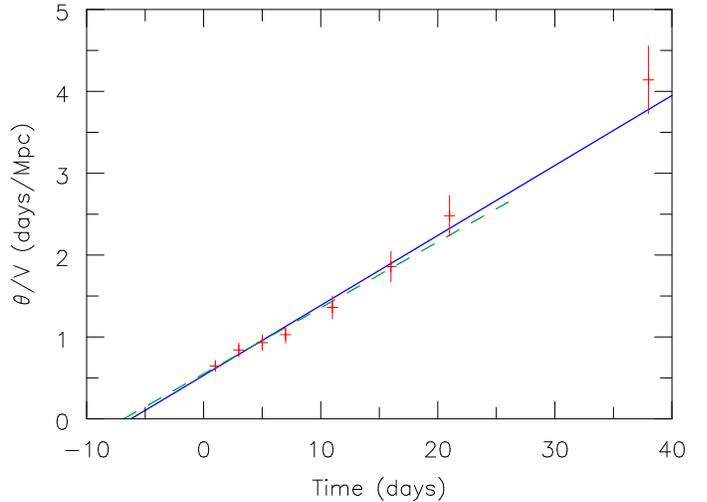}
\caption{
Linear fit to the BV data shown in Table~\ref{tab_epm4} for 8 data
points (solid blue line) and 7 data points (broken green line). 
The reciprocal of the slope is the distance to the SN. The time is measured with
respect to JD 2451480.9. The vertical bars indicate $\pm 10$\%.
(This figure is available in color in the electronic version.)
}
\label{fig_epm4_scatter}
\end{figure}

We now compute the EPM-distance using the model color-temperatures.
First, we use the $\xi$-prescription from Paper II and show resulting EPM quantities
in Table~\ref{tab_epm3}.  Distances derived using the first 7 dates are systematically 
higher than the Cepheid distance, but have large errors. 
The explosion dates, within errors, are consistent. With 8 dates, all distances 
agree within 1 sigma of the Cepheid distance
of 11.7\,Mpc (L03) and the SEAM distance of 12.4\,Mpc (B04, see next section).
Because the observational and model color temperatures show filter set and
epoch dependent differences, the changes in the derived distances between
those shown in Table~\ref{tab_epm2} and Table~\ref{tab_epm3} are
not systematic or uniform.

Finally, combining both color-temperatures and correction factors from the models of
Sect.~\ref{Sec_mod_analysis}, we find even better agreement in the distance predicted between
the different band passes, of 11.7\,Mpc, 11.6\,Mpc, and 11.5\,Mpc, in the usual sequence 
(Table~\ref{tab_epm4}). This is expected since the same models are used ---
indeed the differences, typically less than 2\%, must simply reflect errors in
interpolations etc arising in the application of the EPM technique.
The corresponding explosion dates are -6.2\,days, for all three data sets.
In Fig.~\ref{fig_epm4_scatter}, we show a linear fit of the form $\theta/v_{\rm phot}
= (t-t_{\rm exp})/D$ 
to the BV data shown in Table~\ref{tab_epm4}.

\subsection{Inference based on CMFGEN synthetic SEDs}
\label{Sec_seam}

%
%
%

The Spectral-fitting Expanding Atmosphere Method (SEAM; B04; Mitchell et al. 2002) 
is an attractive alternative to the EPM: each observed SED is directly fitted with a
synthetic SED, produced by a detailed model atmosphere calculation
for the corresponding date, avoiding necessary corrections for approximating
observed colors with those of a single-temperature blackbody (EPM; see 
previous section).
 
Nonetheless, the SEAM follows a similar procedure to the EPM.
Indeed we would expect to derive very similar answers to that found
using the EPM technique with detailed SN fitting.
We first select a number of epochs for which photometric
and spectroscopic observations are available and accurate, and for which
we possess high quality fits to the observed SED.
We then select a range of explosion times, and compute the
photospheric radius expected for each date of observation, given by
$R_{\rm phot}(t) = v_{\rm phot}(t) \, (t-t_{\rm exp})$, where $t$ is the observation date,
$t_{\rm exp}$ the explosion date, and $v_{\rm phot}(t)$ represents the corresponding photospheric
velocity (the initial radius of each considered mass shell is neglected).
As in the EPM, the assumption of homologous expansion, warranted past $\approx$1 day after 
core-bounce, is essential. 

In our approach, we do not evolve an original hydrodynamical input for the exploding envelope;
we use instead an analytical description of the density distribution, and adjust the luminosity 
to match typical values expected from explosion models and former spectroscopic studies
(for SN1999em, we use the Cepheid-distance to ensure our adopted luminosities
are compatible with observed fluxes and magnitudes).
For a chosen $t_{\rm exp}$ corresponds an $R_{\rm phot}(t)$, usually different from that of
the fitting model for that date; we thus scale the model luminosity in proportion with the
square of the change in $R_{\rm phot}(t)$, and update synthetic magnitudes accordingly.
Then, for each choice of the explosion date $t_{\rm exp}$ and for each
observation epoch $t$, we compute the distance modulus
$\mu_S(t_{\rm exp},t)$ for all bandpass sets $S=\{B,V\}$, $\{B,V,I\}$, and
$\{V,I\}$, given by $\mu_S = \langle \mu_X \rangle_S$, where   $\mu_X = m_X -
M_X - A_X$; $X$ represents one of the band passes in the set $S$; $m_X$,
$M_X$, and $A_X$ are the apparent, absolute, and extinction magnitudes in bandpass $X$.
Thus, for each chosen explosion date, we compute, for each set $S$, a distance modulus 
$\mu_S(t_{\rm exp},t)$ averaged over all observation epochs.
The distance modulus selected at the end corresponds to the $t_{\rm exp}$ that
minimizes the scatter of $\mu_S(t_{\rm exp},t)$ over all epochs.


We show results for this sample of observations in Table~\ref{tab_seam}.
The results are grouped by triads.
Using the adopted reddening $E(B-V) = 0.1$ and only seven (eight) observations,
we obtain a distance of $\sim$12.15\,Mpc (11.5\,Mpc), in good agreement with the SEAM-distance 
of B04 (12.4\,$\pm$\,2.5\,Mpc) or the Cepheid distance of Leonard et al. 
(2003; 11.7\,$\pm$\,1\,Mpc).
Limiting the sample to the first four epochs only leads to a significant increase in 
the distance, to $\sim$15.2\,Mpc: the much reduced time-span makes this prediction 
unreliable.
Note also that the use of the $Z-$band has only a modest impact on the resulting distance.

If we assume that our adopted models, which fit well the observations with the choice
of  $E(B-V)$ = 0.1, would still fit the observations adequately with a modulation of
the reddening by $\pm$\,0.05 about this nominal value (see the discussion in Sect.~\ref{Sec_teff_red}), 
then such a reddening uncertainty introduces a distance uncertainty of $\approx \pm$\,0.8\,Mpc 
(equivalent to a $\pm$\,0.15 range in distance modulus). However, 
this represents an upper limit since new models should be computed that fit better
the observations with the newly adopted reddenings. When this is done, the
change in derived temperatures and radii partially compensate for the
change in reddening (Appendix A).
%
%
%

We find that a $\pm$\,10\% error in $v_{\rm phot}$ has a negligible impact 
on the inferred time of explosion (constrained by the {\it relative} evolution of 
magnitudes between different epochs) but affects noticeably the inferred distance:
at a given epoch, smaller (larger) photospheric velocities mean smaller (larger) 
photospheric radii and thus smaller (larger) distances to match the observed magnitudes.
Thus, by itself, the error on $v_{\rm phot}$ (Sect.~\ref{Sec_phot_vel}) introduces an 
uncertainty of $\pm$\,1\,Mpc in the inferred distance.
Finally, we have added an eighth observation to our sample used so
far (that of the 5th of December 1999).
The average distance for the three bands used is then reduced by
$\sim$0.5-1\,Mpc compared to that of the reference case.

\begin{table}
\caption[]
{
SEAM results for the explosion date and distance to SN1999em, under
a variety of choices that highlight uncertainties: bandpass set, 
number of observation dates, reddening, and photospheric velocity values.
Time is relative to JD 2,451,480.54.
}
\label{tab_seam}
\begin{tabular}{ccccccc}
\hline
$t_{\rm exp}$ &   $\mu$  &   $d$  & $E_{B-V}$ &  $v_{\rm scale}$ & $S$  & $n_{\rm obs}$ \\
\hline
(days)&          &  (Mpc) &          &  ($v_{\rm phot}$)&      &               \\
\hline
-6.7  &   30.44  & 12.22  &0.10      &     1.0          &   $BV$ &     7 \\
-6.9  &   30.44  & 12.26  &0.10      &     1.0          &  $BVI$ &     7 \\
-6.8  &   30.41  & 12.10  &0.10      &     1.0          &   $VI$ &     7 \\
\hline
-9.5  &   30.94  & 15.45  &0.10      &     1.0          & $BVIZ$ &     4 \\
-9.1  &   30.90  & 15.14  &0.10      &     1.0          &   $VZ$ &     4 \\
-9.3  &   30.91  & 15.20  &0.10      &     1.0          &  $BVI$ &     4 \\
\hline
-6.9  &   30.44  & 12.26  &0.10      &     1.0          &  $BVI$ &     7 \\
-6.8  &   30.58  & 13.05  &0.05      &     1.0          &  $BVI$ &     7 \\
-6.8  &   30.29  & 11.41  &0.15      &     1.0          &  $BVI$ &     7 \\
\hline
-6.9  &   30.44  & 12.26  &0.10      &     1.0          &  $BVI$ &     7 \\
-7.0  &   30.66  & 13.57  &0.10      &     1.1          &  $BVI$ &     7 \\
-6.9  &   30.21  & 11.02  &0.10      &     0.9          &  $BVI$ &     7 \\
\hline
-6.3  &   30.34  & 11.68  &0.10      &     1.0          &   $BV$ &     8 \\
-6.2  &   30.30  & 11.51  &0.10      &     1.0          &  $BVI$ &     8 \\
-6.0  &   30.25  & 11.22  &0.10      &     1.0          &   $VI$ &     8 \\
\hline
\end{tabular}
\end{table}

\section{Conclusion}
\label{Sec_conc}

In this paper, we have used the model atmosphere code CMFGEN (Paper I and II) 
to perform a quantitative spectroscopic analysis of the photospheric-phase 
evolution of SN1999em, a type II supernovae.

We obtain high-quality fits to flux-calibrated data for eight epochs
(from the 30th of October until the 5th of December 1999), modeling the evolution of the 
optically-thick outflow from full-ionization to recombination.
Indeed, we find that the spectroscopic changes are primarily due to the 
temperature decrease following expansion and radiation-leakage, associated 
with a decrease in the outflow ionization.
The critical shift from Fe{\,\sc iii} to Fe{\,\sc ii} at $T_{\rm phot} \sim 8000$\,K, 
concomitant with the onset of hydrogen recombination, causes a dramatic increase in 
opacity, first in the UV and later in the optical, that alters considerably the 
overall shape of the spectral energy distribution.
For fully or partially ionized models, we were able to reproduce, with good 
accuracy, the shape of both the spectral energy distribution and line profiles.
The thus-constrained chemical composition is compatible with CNO-cycle equilibrium 
abundance values for the first month. 
At early times, the carbon and oxygen abundances are poorly constrained 
due to the lack of lines of C{\sc ii} and O{\sc ii} in the optical and near-IR.
At later times, with further cooling of the outflow, C{\,\sc i} and O{\,\sc i} lines
are seen in the 7000-12\,000\AA\, regions, requiring significant enhancements over the
CNO-cycle equilibrium values; it is not clear if this corresponds to a genuine increase
in the corresponding abundances as the photosphere recedes into deeper layers, or
if this abundance pattern was identical at all previous times.

Significantly, there was no observational incentive to vary the 
metal composition,
whatever the epoch, so we adopted a unique mixture corresponding to solar.
In other words, over a 38-day period, we see no spectroscopic evidence of products from 
explosive-nucleosynthesis.
Additionally, we find no evidence for a flatter density distribution at later times:
for all epochs, we adopt a density exponent $n  = 10-12$ (Paper I-II) which permits 
satisfactory fits to line profile shapes.
Finally, as for other groups (e.g. Mitchell et al. 2002), we have difficulty in reproducing
Balmer (and Paschen, when available) line profiles, both in strength and width.
This discrepancy might stem from a missing outflow energy source, perhaps, e.g., in the form
of energy deposition from radioactive decay of $^{56}$Ni (Mitchell et al. 2002),
or result from the neglect of time-dependence effects (see, e.g., Utrobin \& Chugai 2005)
in standard radiative transfer modeling of photospheric phase type II SN.
We are currently studying this issue.

We also estimate the distance to SN1999em using a variety of approaches and compare
with the reference value of Leonard et al. (2003) using NGC\,1637 Cepheids.
Although we reproduce the under-estimated EPM distance of 
Leonard et al. (2002) or Hamuy et al. (2001), there are non-trivial differences between
our approach and assumptions, and these two works. 
First, our adopted photospheric velocities are higher or lower than theirs, depending on 
the epoch: systematically higher photospheric velocities lead to faster increasing photospheres, which
under the same conditions of magnitude etc..., require a higher distance - the difficulty here is that
the difference in velocity is positive or negative at different dates.
Second, the latest epoch is limited to day 21  (7th observation in our sample) or day 38 (8th observation) 
after discovery - add $\sim$6.2 days to time of core-collapse - in our approach rather than $\ga$ 70days: 
there is a sensible incentive in lengthening
the time baseline to reduce the errors (see Appendix), but as argued in Paper II, the corrupting effect 
of lines in all optical band passes as well as the likely energy contribution from radioactive decay
make such late time observations less adequate for the EPM. Third, to extract the distance and time of explosion
from our collection of $\{t,\theta/v\}$ data points, we use time as the independent variable, rather than 
$\theta/v$: this seems a better approach since time is accurately known. Moreover, in our 
least-square fitting, we adopt constant fractional errors in $\theta/v$  rather than constant weights.
We find that this approach yields distances that are systematically and at least 10\% greater that those obtained
by assuming $\theta/v$ as the independent variable and constant weights.

Applying the correction-factor recipe of Paper II leads to a higher distance, in good agreement
with the Cepheid estimate. It thus seems that correction factors are indeed a major uncertainty in the EPM,
the systematically lower values used previously playing a central role in the discrepancy (see, e.g., Baron et al. 2004).
Overall, however, approaches based on tabulated correction factors lead to a significant disparity in
distances obtained from different bandpass sets, in conflict with the near-constancy, at early times, of 
the angular size of the photosphere with optical-wavelength.
The problem likely stems from the intrinsic scatter of correction factors at a given color-temperature,
obtained from observations alone or even adopted from tailored models.  

When both color-temperatures {\it and} correction factors are adopted from individual best-fit models to each 
observation, the EPM technique delivers more consistent distance estimates between band passes.
Slight differences persist from small but systematic discrepancies between the model and observed
spectral distributions.
But with such an approach, we obtain an EPM-distance of 11.5$\pm$1\,Mpc, in agreement with the
Cepheid estimate.
Hence, while there is nothing fundamentally wrong with the EPM, the need for detailed model
atmosphere calculations defeats the original purpose of the method.
 

We obtain another distance estimate of 12.2$\pm$2\,Mpc using synthetic 
spectra and magnitudes, minimizing the scatter in individual distance moduli for a range 
of adopted explosion dates.
This approach is similar to the approach of Baron et al. (2004) and delivers 
a similar distance. The inferred explosion date is the 23rd of October 1999 with an
uncertainty of 1 day --- this is approximately 4 days earlier than the 
estimate of Hamuy et al. (2001) but similar to that of Leonard et al. (2002)
and Baron et al. (2004). 

This study shows that there are a variety of reliable ways to determine distances
with early-time photospheric-phase type II SN, with good prospects for cosmology. 
In forthcoming studies, we will reproduce the present analysis with a larger sample
of objects, covering a wider range of distances or redshifts, and thereby attempt to 
provide an alternative, to using type Ia SN, for the cosmological distance scale.

\begin{acknowledgements}

We wish to thank Mario Hamuy for providing some of the spectra used in this study, as
well as assistance with the photometry. We also thank Eddie Baron for providing the
HST data for SN1999em taken on the 5th of November, as well as Doug Leonard for
comments on a draft of this paper.
DJH gratefully acknowledges partial support for this work from NASA-LTSA grant NAG5-8211.
Improved data for C{\,\sc i} was kindly supplied by Bob Kurucz.
\end{acknowledgements}

\appendix
\section{Errors}
\label{Errors}

As an accurate distance is the goal of the EPM technique it
is worth discussing the error sources, and how they effect
the determination of the angular diameter, and hence the
distance, in more detail.
For simplicity we consider the use of two filters only.
Let $F_i$ be the observed flux in bandpass $i$, and
$F^c_i$ the reddening corrected flux. Thus

\begin{equation}
F^c_i = F_i \exp\left(0.921R_i E(i-j)\right)
\end{equation}

\noindent
where $R_i=A_i/E(i-j)$ and $E(i-j)$ is the color excess (e.g., $E(B-V)$). The blackbody color temperature 
is determined from

\begin{equation}
{F_i \over F_j} \exp\left(0.921E(i-j)\right)= \left( {\nu_i \over \nu_j} \right)^3
{ {\exp(u_j)-1} \over {\exp(u_i)-1} }
\label{eqn_tcol}
\end{equation}

\noindent
where $u_i = h\nu_i/KT$. Under the assumption that the errors are small,
we can determine them using standard techniques (Taylor 1997). Differentiating 
Eqn.~\ref{eqn_tcol}
yields
\begin{equation}
   {\partial T \over \partial F_i} = {T \over F_i} \, {1 \over G_i -G_j} 
\end{equation}
and 
\begin{equation}
   {\partial T \over \partial F_j} = -{T \over F_j} \, {1 \over G_i -G_j} 
\end{equation}
where
\begin{equation}
G_i=  {u_i \over {1 -\exp(-u_i)} }
\end{equation}

The expression for the angular diameter, $\theta$ is

\begin{equation}
\theta = \left(F_i/B_i \right)^{1/2} {1 \over {\sqrt{\pi} \xi}}
\end{equation}

To allow for the error in $\xi$ we split it into two parts, a random error
at fixed temperature, $\sigma_{\xi T}$, and an error 
attributed to the error in $T$ which can be attributed to an
error in $F_i$,  $F_j$, and $E(i-j)$. Thus, for example,

$${\partial \xi \over \partial F_i} =  
{\partial \xi \over \partial T} {\partial T \over \partial F_i} $$

\noindent 
and where {$\partial \xi/\partial T$} is determined from a set of models. 

If we assume the errors are independent we obtain
\begin{eqnarray}
\sigma_\theta &  = & {\theta \over 2} \left(
  4 \left({\sigma_{\xi T} \over \xi}\right)^2 + \right.  \nonumber \\ 
& & \left. 0.92^2 \left(R_i -{G_i \over G_i -G_j } -
     {2 \over G_i-G_j } {\partial \ln \xi \over \partial \ln T} \right)^2
\sigma^2_{\hbox{E(i-j)}} + \right. \nonumber  \\
& & \left. \left( {G_j \over {G_i-G_j}}  + {2 \over G_i-G_j } {\partial \ln \xi \over \partial \ln T} \right)^2
\left({\sigma_{\hbox{$F_i$}} \over F_i}\right)^2 \right. +  \nonumber \\ 
& & \left. \left( {G_i \over {G_i-G_j}}  + 
     {2 \over G_i-G_j} {\partial \ln \xi \over \partial \ln T} \right)^2
\left({\sigma_{\hbox{$F_j$}} \over F_j}\right)^2    
\right)^{1/2}
\end{eqnarray}

The value of $\sigma_{\xi T}$ is difficult to determine. An estimate
of its value can easily be determined from a grid of models at fixed
effective temperature encompassing a range of SN type models
(e.g., density exponent). However, there is no guarantee that this
grid is representative of true SN. A further source of error is the
fact that models do not exactly match the observations, and this
will contribute an error greater than those introduced by the dispersion in 
SN properties, This error could, if known, be incorporated into $\sigma_{\xi T}$.

In Table \ref{tab_errors}, we list $G_i/(G_i-G_j)$, $d\ln \theta/d\ln F$, 
$d\ln\theta/dE(B-V)$ for three different passband 
combinations. We ignore the coupling with $\xi$, and assume
that the fractional errors in the two pass bands are identical.
As readily apparent, the fractional error in $\theta$ due to an error in
flux increases with SN temperature --- this is simply
due to the increasing insensitivity of the blackbody slope to temperature as
we approach the Rayleigh-Jeans regime. We also note that
a 1\% error in flux is generally more important than a 0.01
error in E(B-V). This somewhat surprising result occurs since
the effect of the reddening on the temperature estimate, and the direct effect
of reddening on the angular diameter (because of a change in flux) partially
compensate. This has been previously noted by Leonard et al. (2002),
Eastman et al. (1996), and Schmidt et al. (1992).
Not surprisingly, we also see that the BI passband combination,
with its wider wavelength coverage, gives potentially the smallest errors.
Finally we note that in the BV band, the angular diameter is very
sensitive to flux errors. A random error of 3\% in each passband
leads to a 10 to 14\% error in $\theta$.

\begin{table}
\caption[]{Errors in $\theta$ due to flux and reddening errors.}
\label{tab_errors}
\begin{tabular}{cccc}
\hline
$T(K)$  & $G_i/(G_i-G_j)$ & $d\ln\theta/d\ln F$\ & $d\ln\theta/dE(B-V)$ \\
\hline
\multicolumn{4}{c}{BV} \\
\hline
 6000  &    5.18  &    3.33  &   -0.50\\
 8000  &    5.48  &    3.54  &   -0.64\\
10000  &    5.89  &    3.83  &   -0.83\\
12000  &    6.36  &    4.16  &   -1.04\\
14000  &    6.87  &    4.52  &   -1.28\\
\hline
\hline
\multicolumn{4}{c}{VI} \\
\hline
 6000  &    3.60  &    2.22  &   -0.56\\
 8000  &    3.94  &    2.46  &   -0.75\\
10000  &    4.35  &    2.74  &   -0.97\\
12000  &    4.78  &    3.05  &   -1.21\\
14000  &    5.24  &    3.37  &   -1.47\\
\hline
\hline
\multicolumn{4}{c}{BI} \\
\hline
 6000  &    2.40  &    1.39  &   -0.54\\
 8000  &    2.57  &    1.50  &   -0.71\\
10000  &    2.77  &    1.64  &   -0.92\\
12000  &    3.00  &    1.80  &   -1.15\\
14000  &    3.24  &    1.97  &   -1.39\\
\hline\\
\end{tabular}\\
\end{table}

Errors associated with $\xi$ are specific to the EPM technique. However
the other error sources will also apply, in a complicated fashion,
to the SEAM technique. The primary distinction is that the SEAM 
technique uses global fitting, whereas the EPM is sensitive to 
local errors in 2 or 3 pass bands. These errors arise,
for example, from errors in the abundance, the neglect of some
species, from observational errors, and from inadequacies in
the model. 

\section{Distance determination}

The relation between the distance and $\theta$ is

\begin{equation}
{\theta \over v} = { {\left(t-t_{\rm exp}\right)} \over D}
\end{equation}
 
Several different approaches can be used to estimate $D$ and $t_{\rm exp}$.
Because $t$ is known accurately, we must use $t$ as the independent variable.
In standard form the equation is
\begin{equation}
y = a +bt
\end{equation}
If we use least squares, errors for $a$ and $b$ can be readily determined
from the accuracy of the fit. Thus we have $D=1/b$ and
\begin{equation}
\sigma_D/D = \sigma_b/b
\end{equation}
Similarly, $t_{\rm exp}=-a/b$ and
\begin{equation}
\sigma_{t_{\rm exp}} = \left( \sigma^2_a + t^2_{\rm exp} \sigma^2_b - 2 t_{\rm exp} \sigma_{ab}
\right)^{1/2}/b
\end{equation}
where $\sigma_{ab}$ is the covariance of $a$ and $b$.

The errors on $\theta/v$ are difficult to determine accurately. However, in
general we expect the errors in $\theta/v$ to scale with $\theta/v$.
For simplicity, we therefore perform weighted least squares fitting assuming
constant fractional errors. Closer examination of the error analysis
discussed earlier suggests that the fractional errors in $\theta/v$ could
be somewhat larger at earlier epochs, but given the additional influence 
of systematic errors, we have made no effort to take this into account.

We also used robust fitting techniques (see Press et~al. 1986)
with weights, but the answers were generally similar 
(i.e., within 1 sigma) to those obtained with least squares fitting.
As a test, we also tried fitting the data assuming constant errors. However,
this gives too much weight to the data of day 38, and leads to unrealistically
small error bars. 

The errors on $D$ and $t_{\rm exp}$ listed in the tables are from the
fit only, and thus are a MINIMUM estimate of the error. 
No systematic errors, which are difficult to determine,
have been taken into account. Systematic errors will arise from
errors in reddening, and possibly, errors in the model. It is also
likely, for example, that errors in adjacent $\theta/v$ estimates
are correlated, since the spectra are similar, and hence model 
errors are likely to be similar.

\end{document}